\pdfoutput=1
\documentclass[12pt]{article}

\usepackage{amsfonts,amsmath,color,graphicx,MnSymbol,esint}
\usepackage{url}
\usepackage{color}
\usepackage{cite}
\usepackage[latin1]{inputenc}

\usepackage[OT2,T1]{fontenc}
\newcommand\textcyr[1]{{\fontencoding{OT2}\fontfamily{wncyr}\selectfont #1}}
\newcommand{\cyrLL}{\text{\textcyr{L}}}

\newcommand\encadremath[1]{\vbox{\hrule\hbox{\vrule\kern8pt 
\vbox{\kern8pt \hbox{$\displaystyle #1$}\kern8pt} 
\kern8pt\vrule}\hrule}}
\def\enca#1{\vbox{\hrule\hbox{
\vrule\kern8pt\vbox{\kern8pt \hbox{$\displaystyle #1$}
\kern8pt} \kern8pt\vrule}\hrule}}

\newcommand\figureframex[3]{
\begin{figure}[bth]
\hrule\hbox{\vrule\kern8pt 
\vbox{\kern8pt \vbox{
\begin{center}
{\mbox{\epsfxsize=#1.truecm\epsfbox{#2}}}
\end{center}
\caption{#3}
}\kern8pt} 
\kern8pt\vrule}\hrule
\end{figure}
}
\newcommand\figureframey[3]{
\begin{figure}[bth]
\hrule\hbox{\vrule\kern8pt 
\vbox{\kern8pt \vbox{
\begin{center}
{\mbox{\epsfysize=#1.truecm\epsfbox{#2}}}
\end{center}
\caption{#3}
}\kern8pt} 
\kern8pt\vrule}\hrule
\end{figure}
}

\makeatletter
\@addtoreset{equation}{section}
\makeatother
\newtheorem{theorem}{Theorem}[section]

\newtheorem{remark}{Remark}[section]
\newtheorem{proposition}{Proposition}[section]
\newtheorem{lemma}{Lemma}[section]
\newtheorem{corollary}{Corollary}[section]
\newtheorem{definition}{Definition}[section]
\def\br{\begin{remark}\rm\small}
\def\er{\end{remark}}
\def\bt{\begin{theorem}}
\def\et{\end{theorem}}
\def\bd{\begin{definition}}
\def\ed{\end{definition}}
\def\bp{\begin{proposition}}
\def\ep{\end{proposition}}
\def\bl{\begin{lemma}}
\def\el{\end{lemma}}
\def\bc{\begin{corollary}}
\def\ec{\end{corollary}}
\def\beaq{\begin{eqnarray}}
\def\eeaq{\end{eqnarray}}
\newcommand{\proof}[1]{{\noindent \bf proof:}\par
{#1} $\square$}

\newcommand{\beq}{\begin{equation}}
\newcommand{\eeq}{\end{equation}}
\newcommand{\bea}{\begin{eqnarray}}
\newcommand{\eea}{\end{eqnarray}}

\renewcommand{\and}{{\qquad {\rm and} \qquad}}

\newcommand{\td}[1]{{\tilde{#1}}}

\newcommand{\ii}{{\mathrm{i}}}
\newcommand{\e}{{\,\rm e}\,}

\newcommand{\Pint}{{\int\kern -1.em -\kern-.25em}}

\renewcommand{\Re}{{\mathrm{Re}}}
\renewcommand{\Im}{{\mathrm{Im}}}

\newcommand{\Arg}{{{\rm Arg\,}}}
\newcommand{\calE}{{{\cal E}}}

\begin{document}

\sloppy

\pagestyle{empty}
\hfill IPHT T13/189\\
\indent \hfill CRM-2013-3328
\addtolength{\baselineskip}{0.20\baselineskip}
\baselineskip 16pt 
\begin{center}
\begin{Large}\fontfamily{cmss}
\fontsize{20pt}{30pt}
\selectfont
\medskip
\textbf{Planar maps, circle patterns and 2d gravity}
\end{Large}\\
\bigskip
\bigskip
{\sl François David}\hspace*{0.05cm}${}^1$, {\sl Bertrand\ Eynard}\hspace*{0.05cm}${}^{2,\,3}$\\
\vspace{6pt}

${}^1$ Institut de Physique Théorique,\\
CNRS, URA 2306, F-91191 Gif-sur-Yvette, France\\
CEA, IPhT, F-91191 Gif-sur-Yvette, France\\
\vspace{6pt}
${}^2$
Institut de Physique Théorique,\\
CEA, IPhT, F-91191 Gif-sur-Yvette, France\\
CNRS, URA 2306, F-91191 Gif-sur-Yvette, France\\
\vspace{6pt}
${}^3$ CRM, Centre de Recherche Math\'ematiques, Montr\'eal QC Canada.\\
\vspace{6pt}
\end{center}

\vspace{20pt}
\noindent{\bf Abstract:}
Via circle pattern techniques, random planar triangulations (with angle variables) are mapped onto Delaunay triangulations in the complex plane.
The uniform measure on triangulations is mapped onto a conformally invariant spatial point process. We show that this measure can be expressed as:
(1) a sum over 3-spanning-trees partitions of the edges of the Delaunay triangulations; 
(2) the volume form of a Kähler metric over the space of Delaunay triangulations, whose prepotential has a simple formulation in term of ideal tessellations of the 3d hyperbolic space $\mathbb{H}_3$;
(3)  a discretized version (involving  finite difference complex derivative operators $\nabla,\bar\nabla$) of Polyakov's conformal Fadeev-Popov determinant in 2d gravity;
(4) a combination of Chern classes, thus also establishing a link with topological 2d gravity.

\vspace{26pt}
\begin{center}
July 11, 2013\\
(misprints corrected) December 19, 2013
\end{center}
\pagestyle{plain}
\setcounter{page}{1}

\newpage
\tableofcontents
\newpage

\section{Introduction}
\newcommand{\EofT}{\mathcal{E}(T)}
\newcommand{\VofT}{\mathcal{V}(T)}
\newcommand{\FofT}{\mathcal{F}(T)}

It has been argued by physicists
\cite{David1984}\cite{Frohlich1985}\cite{Kazakov1985}
, that the continuous limit of large 2-dimensional maps, should be the same thing as the so-called 2d-quantum gravity, i.e. a theory of random surfaces, or also a theory of random metrics (for general references on the subject see e.g. \cite{AmbjornDurhuusJonsson:2005}). 
For 2d quantum gravity, since changing the metrics can be partially absorbed by reparametrizing the surface, Polyakov first proposed to gauge out the diffeomorphism group, by chosing a conformal metrics \cite{Polyakov1981207}.
The measure
(probability weight)
 of the conformal metrics, is then the Jacobian of the gauge fixing operator, which itself can be written as a Gaussian integral over Fadeev-Popov ghosts, with the  quadratic form encoding diffeomorphisms, i.e. the covariant derivative operators.
Polyakov found that the Faddeev-Popov deteminant 
can be evaluated through its trace anomaly, and is the exponential of the Liouville action times the central charge of ghosts $c_{ghosts}=-26$.
Hence 2d quantum gravity can be treated as a conformal 2d field theory: the Liouville theory, where essentially the metrics is locally the exponential of a Gaussian free field, and is a peculiar case of  (bosonic) string theories, often denoted non-critical strings (for a review see e.g. \cite{Nakayama2004}).
The couplings of the Liouville theory is fixed by the consistency condition that the total conformal anomaly (Liouville + ghosts + additional quantum fields living in the metric) must vanish.
The Liouville conformal field theory is defined in the Euclidian plane (and more generally on any open and closed Riemann surfaces), and is characterized by its correlation functions, which are functions of points in the Euclidian  plane. In particular it is characterized by its short distance behaviors, called OPE="operator product expansion", and by the scaling dimensions of its local operators, encoded into the so called KPZ relations
\cite{KPZ:1988}   \cite{David1988-F}  \cite{David1988-M}\cite{DistlerKawai1989}
.

Let us also mention that another approach to 2d quantum gravity, proposed by Witten \cite{Witten:1990}, claims that the probability weight of random surfaces, being metrics independent, is topological, and is a combination of Chern classes. Kontsevich proved later  \cite{Kontsevich1992} that Witten's topological gravity partition function, indeed coincides with KdV tau function, which was the expected continuous limit of large map obtained through matrix model techniques \cite{BrezinKazakov1990}\cite{DouglasShenker1990}\cite{GrossMigdal1990}.

Going back to the discrete case, planar maps have been studied since decades by combinatorial and random matrix methods, and many explicit results corroborate the equivalence beetween the large map limit and 2d quantum gravity (see e.g. 
\cite{KostovPonsotSerban2004}\cite{ChekhovEynardRibaut2013}
).
It has been shown recently, by combinations of combinatorics and probabilistics methods, that the continuous limit (with the Gromov-Hausdorff distance) of large planar maps equipped with the graph distance, exists, and converges as a metric space, towards the so-called "Brownian map"
\cite{LeGall2013}\cite{Miermont2013}
 (see the references therein for previous works and the relation between labelled trees and planar maps).
The problem which has so far remained elusive, is to compare that limit (in the GH topology) with the Liouville conformal field theory in the plane, and to prove their equivalence, if it exists.
Therefore, a requirement in order to check this claim, is to be able to bijectively embed the planar maps into the Euclidian plane.

Many methods of embedding planar maps into the Euclidian plane 
are available. In particular, for planar triangulations, studies involve
 the "barycentric" (or Tutte) embedding (see e.g.  \cite{AmbjornBarkleyBudd2012} ), 
and the "Regge" embedding (where the curvature is located at vertices only) (see e.g. sect. 6 of \cite{Hamber2009}).
``Circle packing" methods (and their extensions known as "circle pattern") are also available (and largely studied in the mathematical literature, for a a review see \cite{Stephenson:2005fk}) and the use of circle packings has been advocated more recently for the problem considered here
(see e.g. \cite{Benjamini2010}).
These later methods have  the advantages that: (1) they provide quasiconformal mappings, hence a control of the conformal properties of the mapping,  (2) the embedding is obtained by a variational principle \cite{ColinDeV1991}, involving an integrable system, and integrability helps a lot (notice that Liouville theory itself is an integrable theory).

In this article, we consider a very natural extension of the circle packing and circle pattern methods, relying on the patterns of circumcircles of Delaunay triangulations. While the circle pattern methods involve (implicitly or explicitly) assigning fixed "intersection angle variables" to the edges of the map, we use the fact that the whole (moduli ) space of surfaces is obtained by varying these angle variables. We thus 
study how the uniform measure on random planar maps, equipped with the uniform Lebesgue measure on edge angles variables, gets transported by the circle pattern embedding method, to a distribution in the Euclidian plane, and we obtain several new and interesting results.
We first show that we obtain a conformally invariant measure on point distributions (i.e. a conformally invariant spatial point process), with an explicit representation in term of geometrical objects on Delaunay triangulations. 
We then show that this distribution is nothing but the volume form of a Kähler metric form on the space of Delaunay triangulations, and identify its prepotential, which turns out to have a geometrical interpretation in term of 3d hyperbolic geometry (in an analogous but different sense than the variational principles underlying circle patterns and quasiconformal mappings). 
We also show that this Kähler measure can be written as a "discrete Fadeev-Popov" determinant (involving  discrete 
complex derivative operators $\nabla,\bar\nabla$), very similar to Polyakov's conformal Fadeev-Popov determinant, thus establishing a new link between planar maps and 2d gravity.
Finally we show that our measure on planar maps can also be written as a combination of Chern classes, thus also establishing a link with topological 2d gravity.

\section{Presentation of the results}
\subsection{Abstract  Euclidean triangulations and Delaunay triangulations}
Let $T$ denote an abstract triangulation of the Riemann sphere.
$\VofT$, $\EofT$ and $\FofT$ denote respectively the sets of vertices $v$, edges $e$ and faces (triangles)  $f$ of $T$.
Let ${\mathcal{T}}_N$ be the set of all such $T$ with $N=|\VofT|$ vertices, hence $|\EofT|=3(N-2)$ and $|\FofT|=2(N-2)$.

An Euclidean triangulation $\widetilde T=(T,\boldsymbol{\theta})$ is a triangulation $T$ plus an associated edge angle pattern $\boldsymbol{\theta}=\{\theta(e); e\in \EofT\}$ such that the $\theta(e)\in [0,\pi)$.
$\widetilde{\mathcal{T}}_{N}^{f}$ will denote the set of flat Euclidean triangulations, where flat means that for each vertex $v\in \VofT$, the sum of the angles of the adjacent edges satisfy
\begin{equation}
\label{sumthetav}
\sum_{e\to v} \theta(e)=
2\pi
\end{equation}
Note that not all planar triangulations $T\in \mathcal{T}_N$ are present in $\widetilde{\mathcal{T}}_N^{f}$.
However ``generic simple triangulations'' are expected to be in $\tilde{\mathcal{T}}_N$.

\begin{figure}[h]
\label{pointinfty-2} 
\begin{center}
\includegraphics[width=0.9\textwidth]{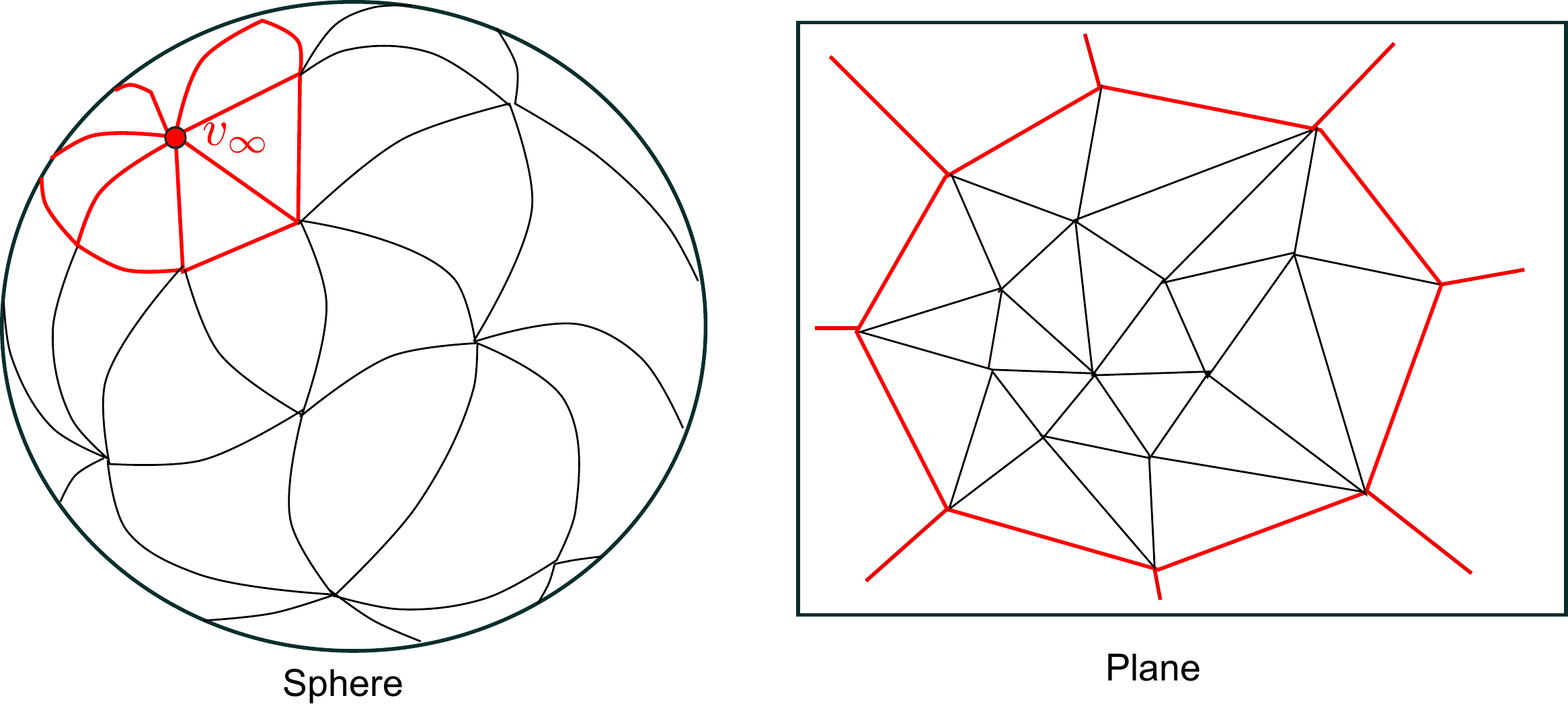}
\caption{
An abstract triangulation $\widetilde{T}$ of the sphere and the corresponding Delaunay triangulation $\mathfrak{T}$ of the plane. The neighborhood of the vertex at $\infty$ is the exterior of the convex hull of  $\mathfrak{T}$.
The circumcircles of the triangles containing $v_\infty$ are straight lines tangent to the boundary of the convex hull.
Requiring that those "circles" meet at given angles $\pi-\theta_e$, fixes the orientation of the half-lines starting from the summits of the convex hull.}
\end{center}
\end{figure}

\begin{figure}[h]
\begin{center}
\includegraphics[width=2in]{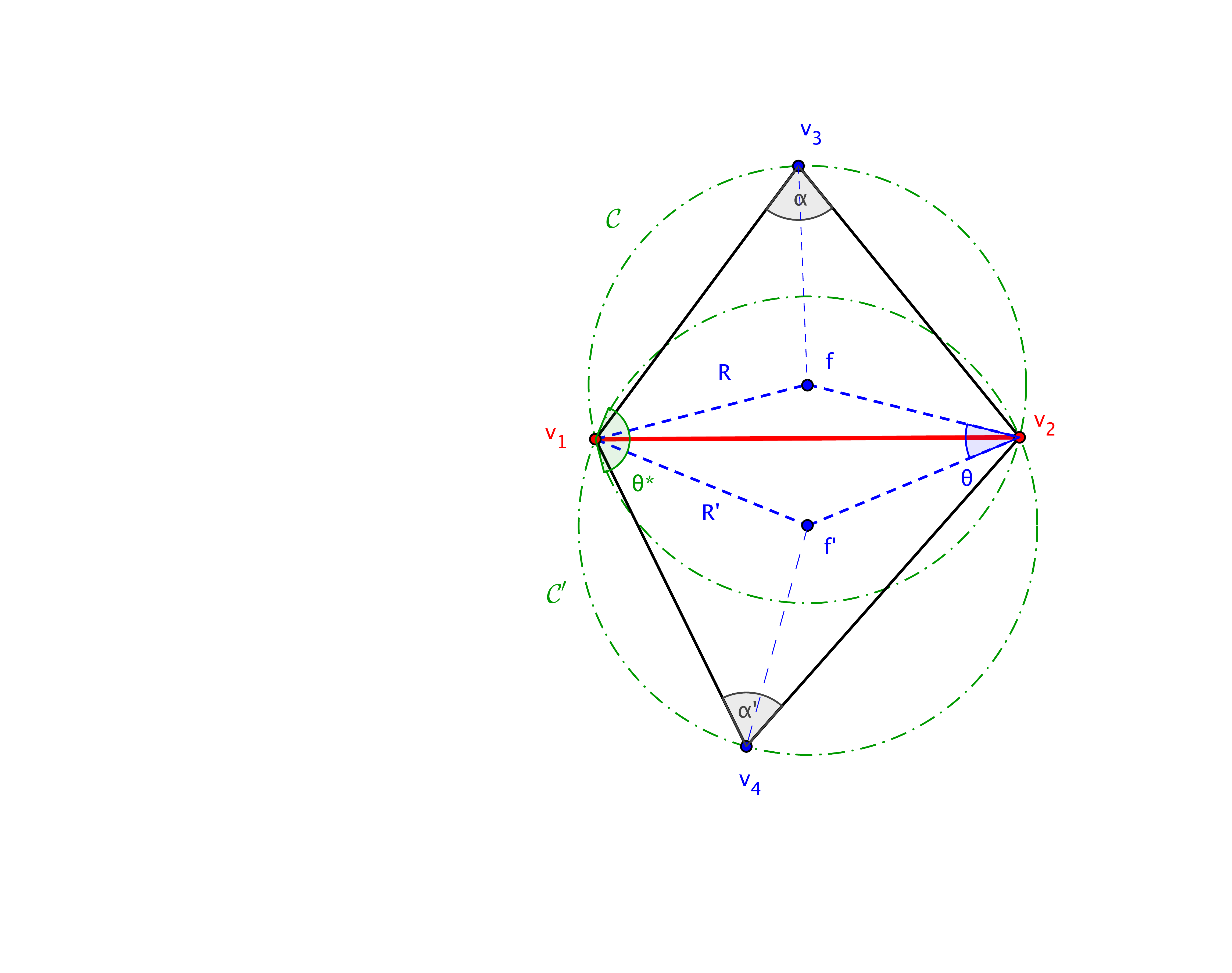}
\caption{The triangles $f$ and $f'$, the circumcircles $\mathcal{C}$ and $\mathcal{C}'$ and angles $\theta$ and $\theta^*=\pi-\theta$ associated to an edge $e=(z_1,z_2)$ of a Delaunay triangulation. $R$ and $R'$ are the radii of the circles $\mathcal{C}$ and $\mathcal{C}'$.}
\label{2triangle-1}
\end{center}
\end{figure}

A theorem by Rivin \cite{Rivin1994} states that there is an angle pattern preserving bijection between $\widetilde{\mathcal{T}}_N^{f}$ and the set of Delaunay triangulations of the complex plane (modulo Möbius transformations) $\boldsymbol{\frak{D}}_N=\mathbb{C}^N/SL(2,\mathbb{C})$.
More precisely, let $Z=\{z_v\}$ be a set of $N$ (distinct) points $v$ with complex coordinate $z_v$ on the complex plane. For simplicity select 3 distinct points ($v_0,v_1,v_\infty)$ and fix through a SL(2,$\mathbb{C}$) transformation their coordinates $(z_{v_0},z_{v_1},z_{v_\infty})$ to $(0,1,\infty)$. To $Z$ is associated the planar Delaunay triangulation $\frak{T}$ by the standard Voronoï/Delaunay construction (the neighbors of $v_\infty$ being on the convex hull of the Delaunay triangulation of $Z\backslash{\{z_{v_\infty}\}}$, see fig.~\ref{pointinfty-2}). 
As depicted on fig.~\ref{2triangle-1}, to an edge $e=(z_1,z_2))$ are associated the upper and lower triangles (faces)  $f=(z_1,z_2,z_3)$ and $f'=(z_2,z_1,z_4)$, and their circumcircles $\mathcal{C}$ and $\mathcal{C}'$. $\theta^*(e)=\alpha+\alpha'$ denotes the intersection angle of the two circles and $\theta(e)=\pi-\theta^*(e)$ is the angle $(f',v_1,f)=(f,v_2,f')$  between a vertex and the center $z_f$ and $z_f'$ of the circumcircles $\mathcal{C}$ and $\mathcal{C}'$ of the faces $f$ and $f'$. A triangulation $\mathfrak{T}$ of the plane is Delaunay if for any face $f=(z,z',z'')$, all the other vertices $z'''$ of $\mathfrak{T}$ are outside  the circumcircle of $f$.

Rivin's theorem states that these is a bijection $\widetilde{T}=(T,\boldsymbol{\theta})\leftrightarrow Z$ 
(with 3 points fixed) 
between $\widetilde{\mathcal{T}}_N^{f}$ and $\boldsymbol{\frak{D}}_N$, which identify the abstract triangulations $T$ and the Delaunay triangulation $\mathfrak{T}$, and the flat edge angle pattern $\theta(e)$ to the circle angle pattern $\theta(e)=\pi-\theta^*(e)$.
It is an extension of the famous theorem by Koebe-Andreev-Thurston  \cite{Koebe1936} stating that there is a bijection between (simple) triangulations and circle packings in complex domains, modulo global conformal transformations, which is widely used to construct quasi-conformal mappings and in conformal geometry.
The proof of Rivin's theorem rely on the same kind of convex minimization functional, using hyperbolic 3-geometry, than for the original circle packing case (see \cite{Rivin1994} and \cite{BobenkoSpringborn2004}).

The concept of Euclidean triangulations can be generalized to non-flat triangulation with defect angles at the vertices
\begin{equation}
\label{sumthetaN}
\sum_{e\to v} \theta(e)=
\Xi(v)\neq
2\pi
\end{equation}
(which can be mapped to Delaunay triangulation of locally flat surfaces with conical singularities with defect angles $2\pi-\Xi(v)$ at the points $z_v$), to triangulations of genus $g>0$ surfaces, in particular periodic triangulations ($g=1$) and to surfaces with boundaries.

\subsection{The measure $\mathcal{D}$ on Delaunay triangulations}
Our purpose is to use this bijection to construct conformally invariant measures on the ensemble $\widetilde{\mathfrak{T}}_N$ of distributions of $N$ points $Z=\{z_1, \cdots z_N\}$ in the complex plane (i.e. spatial point processes on $\mathbb{C}$), via their associated Delaunay triangulation, and to study the properties of these measures. This allows to discuss their relations with the continuum field-theoretic formulation of 2 dimensional quantum gravity and with Liouville theory, as well as their relation with the simpler discrete combinatorial models of random planar maps and matrix models.

Since the SL(2,$\mathbb{C}$) invariance allows to fix 3 points in the triangulations, from now on we work with triangulations and points ensembles with $M=N+3$ points.

For this, we start from the uniform flat measure on $\widetilde{\mathcal{T}}_{N+3}$.

\begin{definition}
We take the measure $\mu(\widetilde{T})=\mu(T,d \boldsymbol{\theta})$ on $\widetilde{\cal T}_{N+3}$ to be the discrete uniform measure on triangulations (as in random planar map ensembles) times the flat Lebesgue measure on the angles $\theta(e)$'s (constrained by \eqref{sumthetav}):
\beq
\label{muT}
\mu(\widetilde{T})=\mu(T,d \boldsymbol{\theta})= {\rm uniform}(T)\ \prod_{e\in \EofT}d\theta(e)\quad \prod_{v\in \VofT}\,\delta\left(\sum_{e\mapsto v} \theta(e)-2\pi\right)
\eeq
\end{definition}
The uniform({$T$}) term in general takes into account the symmetry factor of $T$. If we choose to fix a point $v_\infty$ at infinity, it is always equal to one \footnote{This is analogous to what is done in the theory of random planar maps, i.e. random planar lattices, by dealing with rooted triangulations, i.e. specifying a vertex and an adjacent oriented link}.
It is known that the space $\widetilde{\mathcal{T}}_{N+3}$ is made of the small pieces $\widetilde{\mathcal{T}}(T)$ associated to each triangulation $T$, glued at their boundary where one of the $\theta(e)=0$, and where a flip of the edge $e\in \EofT\leftrightarrow e'\in \EofT'$ occurs. So it is a connected, piecewise linear space. Its dimension as a real manifold (in fact an orbifold) is $2N$.
\begin{figure}[h]
\begin{center}
\vskip -2.em
\includegraphics[height=1.5in]{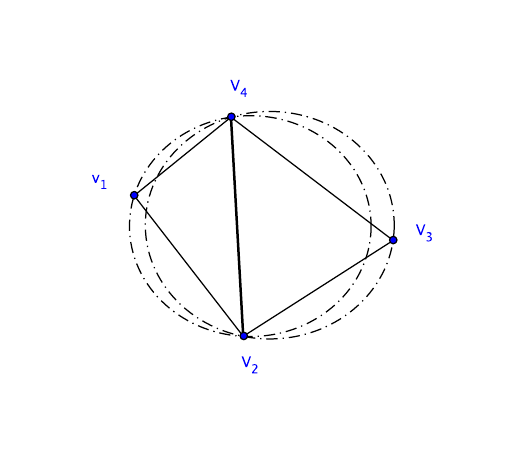}
\hskip -1em\raisebox{10ex}{$\to$}\hskip-1em
\includegraphics[height=1.5in]{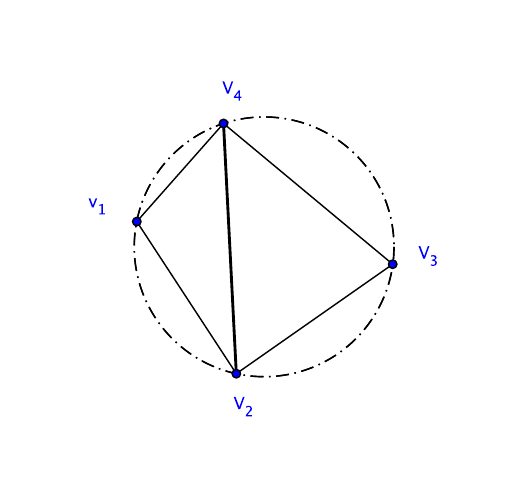}
\hskip -1em\raisebox{10ex}{$\to$}\hskip-1em
\includegraphics[height=1.5in]{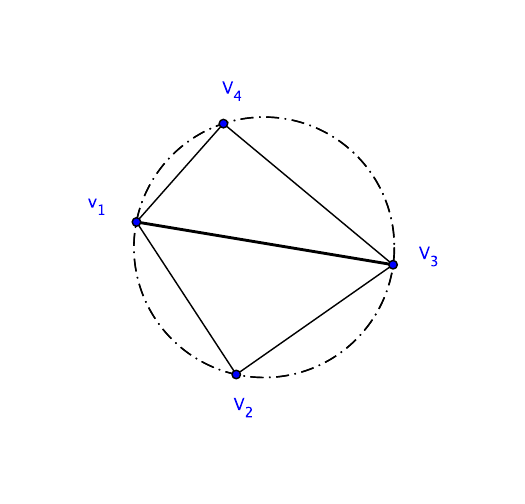}
\hskip -1em\raisebox{10ex}{$\to$}\hskip-1em
\includegraphics[height=1.5in]{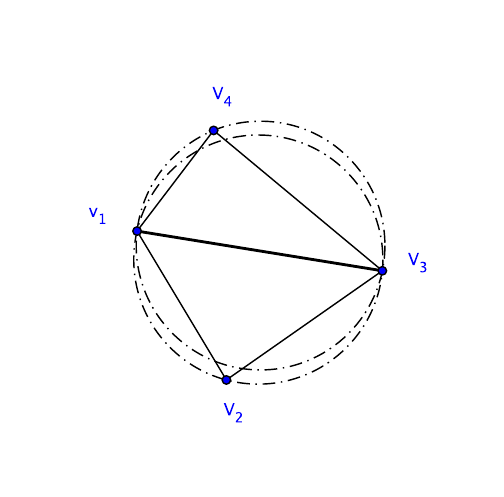}
\vskip -2.em
\caption{The basic edge flip (Whitehead move)  $e\to e'$ between triangulations occurs when $\theta(e)=0$}
\label{FigFlip}
\end{center}
\end{figure}

A first step to study this space is to construct for any triangulation $T$ in $\widetilde{\mathcal{T}}_{N+3}$ a basis $\mathcal{E}_0\subset \EofT$ of $2N$ independent edges, such that given the 2N  angles $\{\theta(e),\,e\in\mathcal{E}_0\}$, the remaining $N+3$ $\{\theta(e'),e'\notin\mathcal{E}_0\}$ can be reconstructed out of the $N+3$ constraints \ref{sumthetav}. 
Such basis are characterized by the following theorem, which is most plausibly already known

\begin{figure}
\begin{center}
\includegraphics[width=3in]{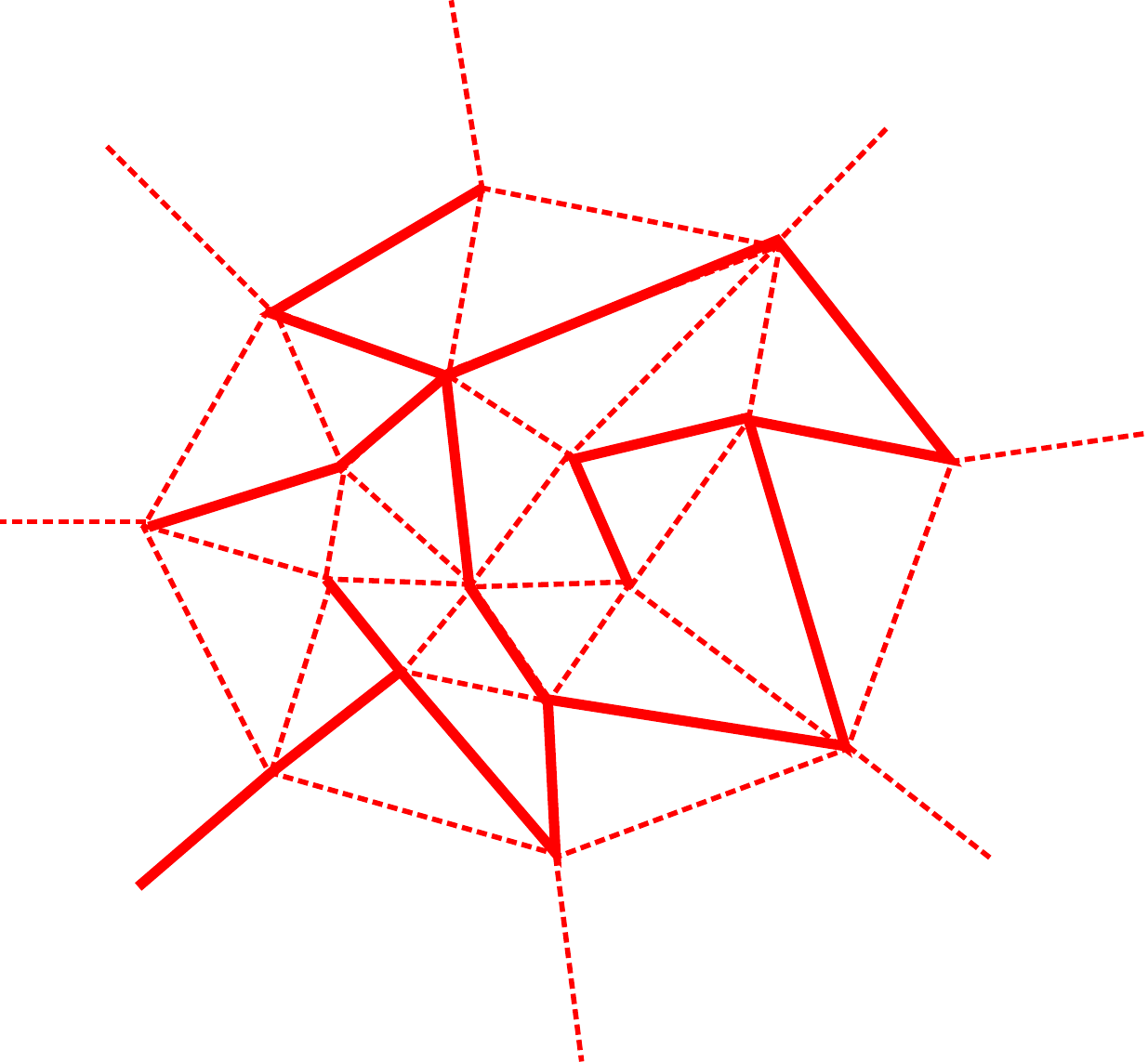}
\caption{A triangulation $T$ (with one point at $\infty$) and an odd CRST  $\bar{\mathcal{E}}_0$  (thick lines).
Its complement $\mathcal{E}_0$ (dashed edges) form an independent basis of angle variables.}
\label{pointinfty-CRST}
\end{center}
\end{figure}

\bt
\label{ThE0def}
A set ${\cal E}_0\subset \EofT $ 
of $2N$ edges, is a basis if and only if its complementary 
$\bar \calE_0 =\EofT\setminus \calE_0$
form a cycle-rooted spanning tree of the triangulation $T$, whose cycle is of odd length.
\et
A cycle-rooted spanning tree (CRST)  $t$ of a graph $\mathcal{G}$ is a connected subgraph of $\mathcal{G}$ which contains all the vertices of $\mathcal{G}$ and has as many edges as vertices. It is composed of a single cycle and of trees attached to this cycle.
See fig.~\ref{pointinfty-CRST}

Choosing a specific basis of independent edges angles $\{\theta(e); e\in\calE_0\}$, i.e. an odd CRST on ${T}$, is not very important. Indeed one has
\bt
\label{theorem2}
The measure on $\widetilde{\cal T}_{N+3}$ can be written
\beq\label{eqmeasuretdTdthetaE0bis}
\mu(T,d\boldsymbol{\theta})\ =\ 
\frac{1}{2} \,\,{\rm uniform}(T) \,\times\,\, \prod_{e\in \calE_0(T)}\,d\theta(e)
\eeq
and is  independent of a choice of basis $\calE_0(T)\subset \calE(T)$ for each triangulation $T$.
\et

We now look at the induced measure on the space $\mathfrak{D}_{N+3}$ of Delaunay triangulations on the plane.
For simplicity, from now on we consider triangulations $T$ with $N+3$ vertices, such that the first three points $(z_1,z_2,z_3)$ are fixed, and can be taken to be the three vertex of a given triangle $f_0$ of $T$. The remaining variables are the complex coordinates $Z_{\setminus\{1,2,3\}}=\mathbf{z}=(z_4,\cdots,z_{N+3})$ of the $N$ free points. 
For a given $\mathbf{z}\in\mathbb{C}^N$ there is generically a single Delaunay triangulation $T$. We shall denote $\theta_e$ the angle associated to the edge $e$ of $T$, and $w_f$ the complex coordinate of the center of the circumcircle of the triangle $f$.

So this measure is simply  given locally by a Jacobian determinant
\begin{equation}
\label{muTz}
\mu(T,d\boldsymbol{\theta})\ =\ d\mu(\mathbf{z})\ =\ \prod_{v=4}^{N+3} d^2 z_v\ \left| 
{\color{black}{   \left({2\over i}\right)^N }}
\det\left( J_T(z)_{\setminus\{1,2,3\}\times {\mathcal{E}}_0}\right)\right|
\end{equation}
where $d^2z=dx\,dy={\ii\over 2}{dz\wedge d\bar z}$ is the flat measure on $\mathbb{C}$, and with $J_T(z)_{\setminus\{1,2,3\}\times {\mathcal{E}}_0}$ the $2N\times 2N$ Jacobian matrix (associated to a basis $\mathcal{E}_0$ of independent edges) obtained from the $(2N+6)\times (3N+3)$ partial derivative matrix $J_T(z)$
\begin{equation}
\label{JTotal}
{J_T}(z)=\left({\partial \theta_e\over\partial (z_v,\bar z_v)}\right)_
{\begin{smallmatrix}
  e\,\in\,\EofT  \\
  v\in\VofT
\end{smallmatrix}
}
\end{equation}
by removing the 6 lines associated to the three fixed vertices $(z_1,\bar z_1,z_2,\bar z_2,z_2,\bar z_3)$, and the 3 columns $(e_1,e_2,e_3)$. From theorem~\ref{theorem2} the Jacobian determinant is independent of the choice of $\mathcal{E}_0$. So
\begin{definition}
The $N$ point density measure associated to the (Delaunay) triangulations $T$ with the 3 fixed points $\{1,2,3\}$ is 
\begin{equation}
\label{defDJacob}
{\mathcal{D}_T(z)}_{\setminus\{1,2,3\}}=\left| 
{\color{black}{    \left({2\over i}\right)^N   }}
\det\left( J_T(z)_{\setminus\{1,2,3\}\times {\mathcal{E}}_0}\right)\right|
\end{equation}
It is a function of the $N+3$ complex coordinates $\textbf{z}=(z_1,\cdots, z_{N+3})$, and is well defined when no points coincide.
\end{definition}

The matrix elements of $J_T(z)$ are easy to calculate. For a given triangulation $T$, the elements 
\begin{equation}
\label{Jelemt}
{J}_{v,e}={\partial \theta_e\over\partial z_v}\,\quad,\qquad {J}_{\bar v,e}={\partial \theta_e\over\partial \bar z_v}
\end{equation}
are non zero only if the vertex $v$ is a vertex of one of the two triangles $f$ and $f'$ that share the edge $e$ (see fig.~\ref{2triangle-1}). With the notations of fig.~\ref{2triangle-1} one has explicitly for the edge $e=(v_1,v_2)$
\begin{align}
\label{JveExpl}
  J_{v_1,e}=  {\ii\over 2}\left({1\over z_{v_4} - z_{v_1} } -{1\over z_{v_3}- z_{v_1} }\right)  &\quad,\qquad  J_{v_2,e}=  {\ii\over 2}\left({1\over z_{v_3} - z_{v_2} } -{1\over z_{v_4}- z_{v_2} }\right) \nonumber \\
   J_{v_3,e}=   {\ii\over 2}\left({1\over z_{v_3} - z_{v_1} } -{1\over z_{v_3}- z_{v_2} }\right)  &\quad,\qquad    J_{v_4,e}=   {\ii\over 2}\left({1\over z_{v_4} - z_{v_2} } -{1\over z_{v_4}- z_{v_1} }\right) 
\end{align}

\subsection{Spanning trees representation of $\mathcal{D}$}

\ref{JveExpl} implies that ${\mathcal{D}_T(z)}_{\setminus\{1,2,3\}}$ is locally a rational function of the $(z_i,\bar z_i)$, and is the determinant of a "derivative-like" operator involving only neighbors vertices and links in the triangulation $T$, like Dirac and Laplace operators. The determinant of Laplace operators on graphs are known to have a representation in term of spanning trees (or extensions). ${\mathcal{D}_T(z)}_{\setminus\{1,2,3\}}$ has also such a representation, but involving more complicated geometrical objects.
\bd[triangle rooted spanning 3-tree]\ \\
\label{defTRS3T}
Let $T$ be a planar triangulation with $N+3$ vertices, and ${\triangle}=f_0$ be a face (triangle) of $T$, with 3 vertices
 $\mathcal{V}({\triangle})=(v_1,v_2,v_3)$ and 3 edges $\mathcal{E}(\triangle)=(e_1,e_2,e_3)$. Let $\EofT_{\setminus\triangle}=\EofT\setminus\mathcal{E}(\triangle)$ be the set of $3N$ edges of $T$ not in $\triangle$.
 
\begin{figure}
\begin{center}
\includegraphics[width=2in]{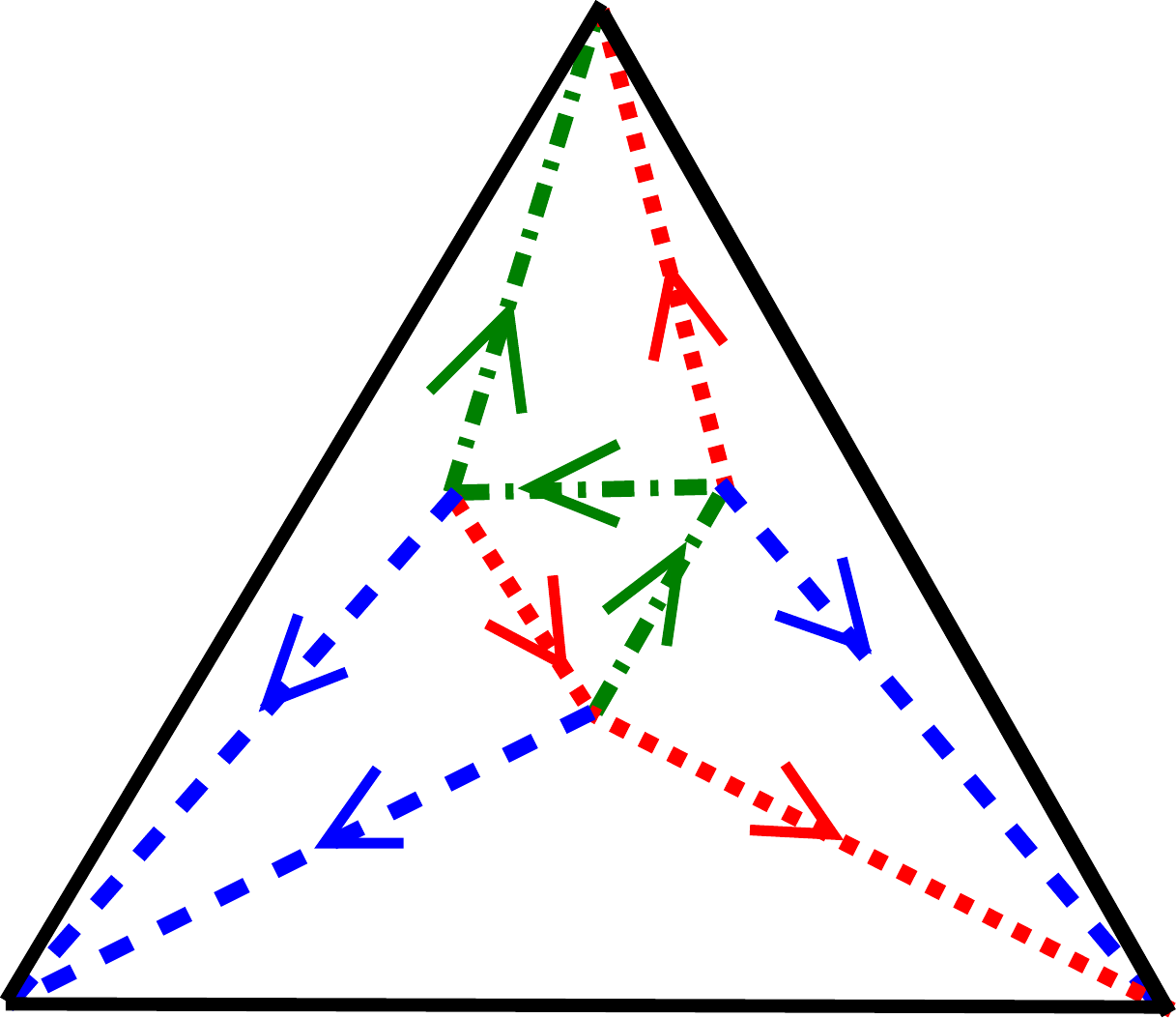}\qquad
\includegraphics[width=2in]{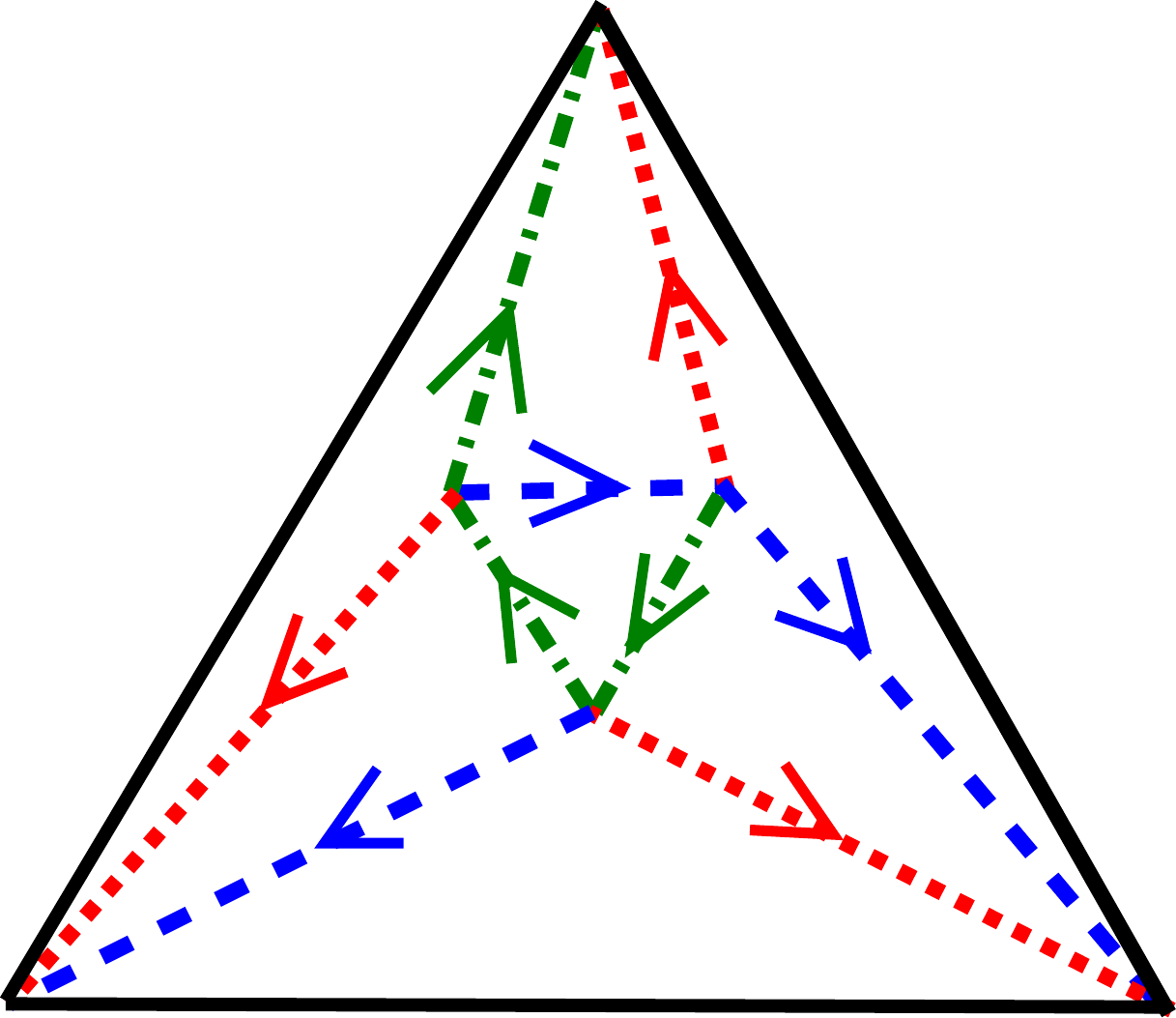}
\caption{2 inequivalent triangle-rooted spanning 3-trees of a $N=3$ planar triangulation (here the octahedron). The fixed triangle $\triangle$ is here the exterior black triangle. The three trees $\mathcal{I}$, $\mathcal{I}'$ and $\mathcal{I}''$ are respectively made of the dotted-red, dashed-blue and dot-dashed-green edges. The natural orientation towards the triangle $\triangle$ is depicted.}
\label{TR3Tree}
\end{center}
\end{figure}

 We call a $\triangle$-rooted 3-tree of $T$ ($\triangle\! R3T$) a family $\mathcal{F}$ of three \emph{disjoint} subsets $(\mathcal{I},\mathcal{I}',\mathcal{I}'')$ of edges of $\EofT$ 
 such that:
 \begin{enumerate}
  \item $(\mathcal{I},\mathcal{I}',\mathcal{I}'')$ are disjoint and disjoint of $\mathcal{E}(\triangle)$
  \item Each $\mathcal{I}\cup\mathcal{E}(\triangle)$, $\mathcal{I}'\cup\mathcal{E}(\triangle)$, $\mathcal{I}''\cup\mathcal{E}(\triangle)$ is a cycle rooted spanning tree of $T$ with cycle $\triangle$.
\end{enumerate}
\ed
 It follows that each $\mathcal{I},\,\mathcal{I}',\,\mathcal{I}''$ contains $N$ edges. There is a natural orientation $e\to \vec e$ of their edges, that we take to be ``pointing towards the triangle $\triangle$''.
 See fig.~\ref{TR3Tree} for a simple illustration.
 NB: the 3-trees defined here look like but are \emph{not} Schnyder woods! \cite{Schnyder1989}

When we keep the vertices $(v_1,v_2,v_3)$ of the triangle $\triangle$ fixed, the measure determinant $\mathcal{D}_T(z)_{\setminus\{1,2,3\})}$ defined in \ref{defDJacob} can be rewritten as a sum over the $\triangle$-rooted  3-trees of $T$ of products of the building blocks  $J_{v,e}$ and $\bar J_{v,e}$ of the Jacobian matrix $J$. More precisely
\bt
\label{ThDas3T}
Let $T$ be a planar triangulation of the plane with $N+3$ vertices. If the 3 fixed points $(v_1,v_2,v_3)$ belong to a triangle (face) of $T$, the measure determinant takes the form
\begin{align}
 \mathcal{D}_T(z)_{\setminus\{1,2,3\})}\ =   &\  
{ \color{black}{\left({1\over 2 \,\ii}\right)^N}}
 \ \sum_{
 \begin{smallmatrix}
 \mathcal{F}=(\mathcal{I},\mathcal{I}'\!,\mathcal{I}'')\\
\triangle\!R3T\,\text{of}\ T
 \end{smallmatrix}}  
 \epsilon(\mathcal{F})\ 
 \times
 \prod_{\vec e=(v\to v')\in\,\mathcal{I}} {1\over z_v-z_{v'}}
 \times
 \prod_{\vec e=(v\to v')\in\,\mathcal{I}'} {1\over \bar z_v-\bar z_{v'}} 
 \label{Das3trees}
\end{align}
where $\epsilon(\mathcal{F}) = \pm 1$ is a sign factor, coming from the topology of $T$ and of $\mathcal{F}$, that is defined explicitly by Eq.~\ref{Fexpl}.
\et
The proof is given in Sec.~\ref{proofDas3T}.
This theorem is non trivial, and is not obtained simply by the Cauchy-Binet expansion of the determinant of $\mathcal{D}_T(z)$ in terms of permutations. Its proof involves also techniques developed in the derivation of the very nice and useful representation of $\mathcal{D}$, that we present in Th.~\ref{TDasKha}.

Remark: The trees $\mathcal{I}$, $\mathcal{I}'$ and $\mathcal{I}''$  play equivalent roles, but only the edges of the first two appear in the r.h.s. of 
\ref{Das3trees}.  Each term in the sum is complex, but exchanging $\mathcal{I}$ and $\mathcal{I}'$ exchanges the $z_v$'s and the $\bar z_v$'s, and in fact does not changes the sign of the $\epsilon(\mathcal{F})$ prefactor, so that the determinant is real.

\ref{Das3trees} shows that $\mathcal{D}_T(z)$ diverge when two or more points $z_v$ coincide. However it allows to control the singularity.

\begin{theorem}
\label{ThConvD}
Each individual term in the sum of \ref{Das3trees} is singular when some $z_v$'s coincide, but the associated (complex) measure
\begin{equation}
\label{dmuF}
d\mu_\mathcal{F}(z)= \prod_{v\notin\triangle} d^2 z_v\  \prod_{\vec e=(v\to v')\in\,\mathcal{I}} {1\over z_v-z_{v'}}
 \times
 \prod_{\vec e=(v\to v')\in\,\mathcal{I}'} {1\over \bar z_v-\bar z_{v'}} 
\end{equation}
is well defined and absolutely integrable.
\end{theorem}
The proof is discussed in Sec.~\ref{proofThConvD}. The fact that the total determinant $\mathcal{D}(z)$ defines an integrable measure is not surprising, since the original flat measure $\mu(T,\boldsymbol{\theta})$ over the angles $\theta(e)$ is finite. The decomposition \ref{Das3trees} is interesting since it takes into account many cancellations in the expansion of the determinant, so that each term is also integrable, and can be used to study the properties of the measure.

\subsection{$\mathcal{D}$ as a Kähler volume form and hyperbolic geometry}
Here we show that the space $\boldsymbol{\mathfrak{D}}_{N+3}\simeq \mathbb{C}^N$ of Delaunay triangulations with 3 point fixed is embodied with a natural structure of a $N$-dimensional Kähler manifold, and that the measure determinant $\mathcal{D}(z)$ is nothing but the volume form for this Kähler metric.

\begin{figure}[h]
\begin{center}
\includegraphics[width=5in]{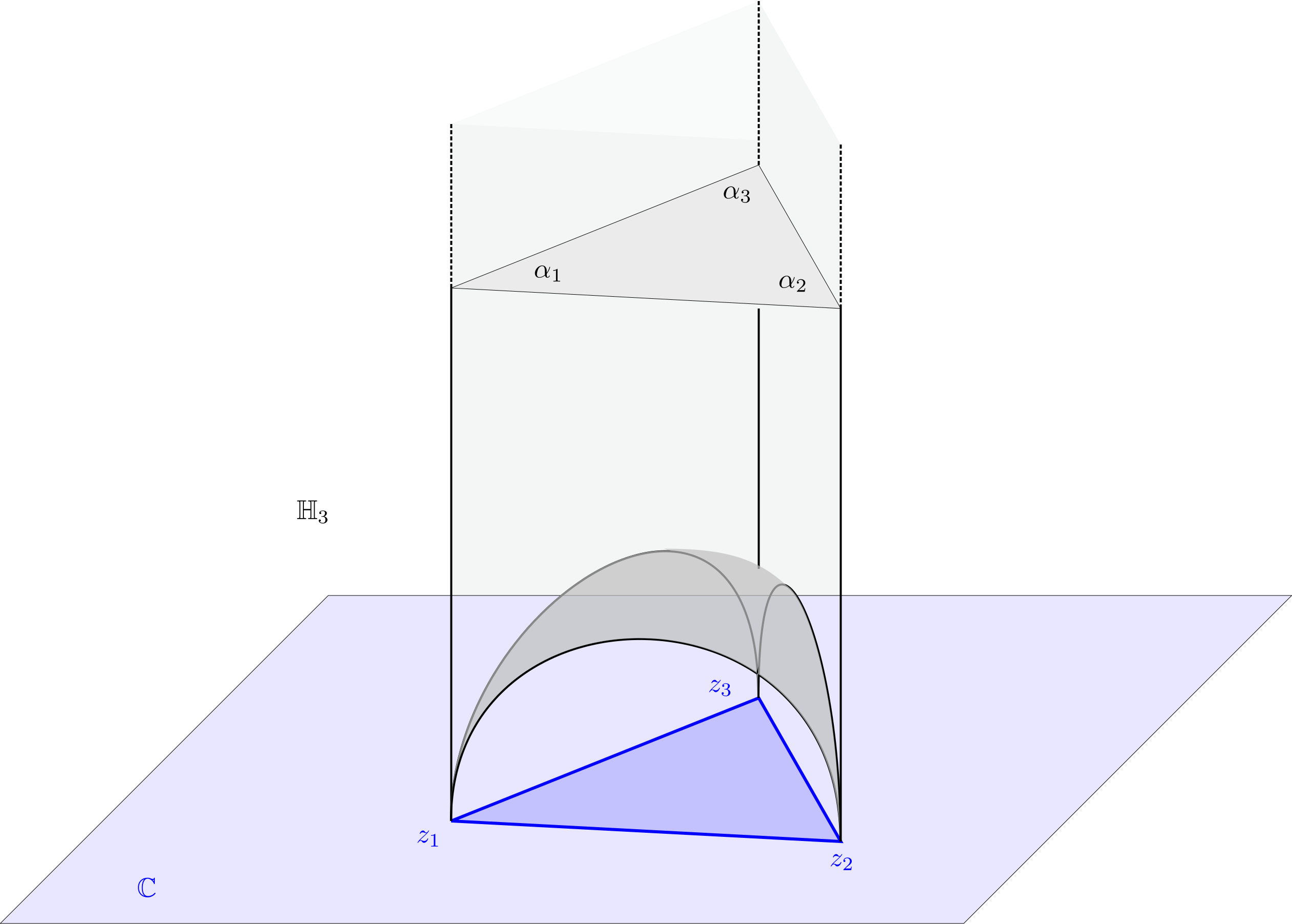}
\caption{A triangle $f=(z_1,z_2,z_3)$ in the plane and the associated ideal tetrahedron in the upper-space $\mathbb{H}_3$.}
\label{fIdealT}
\end{center}
\end{figure}

In the geometry of circle packings and circle patterns in 2 dimensions, 3 dimensional hyperbolic geometry plays a central role. Let us recall a few definitions and basic facts. The complex plane $\mathbb{C}$ is considered as the boundary of the 3 dimensional hyperbolic space $\mathbb{H}_3$, represented by the upper half-space above $\mathbb{C}$. Adding the point at $\infty$, the Riemann sphere $\mathcal{S}_2$ is therefore the boundary of the 3 dimensional Poincaré ball $\mathbb{B}_3$. 
\begin{definition}[Hyperbolic volume of a triangle]
Let $f=(z_1,z_2,z_3)$ be a (oriented) triangle in $\mathbb{C}$. The hyperbolic volume $\mathrm{Vol}(f)$ of the triangle $f$ is the hyperbolic volume of the ideal tetrahedron in $\mathbb{H}_3$ with vertices $(z_1,z_2,z_3,\infty)$ on its boundary.
In term of the coordinates it is given by the Bloch-Wigner function
\begin{equation}
\label{VolHLi}
\mathrm{Vol}(f)= \Im(\mathrm{Li}_2(z))+\ln(|z|) \Arg(1-z)
\quad\text{with}\quad z={z_3-z_1\over  z_2-z_1}
\end{equation}
where $\mathrm{Li}_2$ is the dilogarithm function.
In term of the angles $(\alpha_1,\alpha_2,\alpha_3)$ of the triangle it is given by
\begin{equation}
\label{VolHd}
\mathrm{Vol}(f)=\cyrLL(\alpha_1)+\cyrLL(\alpha_2)+\cyrLL(\alpha_3)
\end{equation}
where \cyrLL \  is the Lobatchevski-Milnor function
\begin{equation}
\label{LobL}
\cyrLL(\alpha)=-\int_0^\alpha d\theta\ \log(2 \sin(\theta))
\end{equation}
\end{definition}
See fig.~\ref{fIdealT}. Vol($f$) satisfies the basic variation formula ($\ell_i$ being the length of the side of the triangle opposite to the angle $i$.
\begin{equation}
\label{dVolH}
d\mathrm{Vol}(f)=-\sum_{\text{3 angles}} d\alpha_i\ \ln(\sin(\alpha_i))\ =\ -\sum_{\text{3 angles}} d\alpha_i\ \ln(\ell_i))
\end{equation}

Now let us consider a planar Euclidean triangulation $\tilde T$ on the sphere and the associated Delaunay triangulation $T$ in $\mathbb{C}$, where 3 points are fixed. We define an action (or prepotential) $\mathcal{A}_T$ for $T$ simply as
\begin{definition}[Action-prepotential of a triangulation]
The action $\mathcal{A}_T$ of a Delaunay triangulation $T$ is defined as {\bf{minus}} the sum of the hyperbolic volumes of its faces (triangles)
\begin{equation}
\label{ATsumV}
\mathcal{A}_T\ =\ -\sum_{\text{triangles}\ f\in\FofT} \mathrm{Vol}(f)
\end{equation}
\end{definition}
\paragraph{Remark:} Some care should be exercised in this definition with the 3 fixed points and the point at $\infty$. Here we shall consider two special cases:
\begin{description}
  \item[Point at infinity:]
  \label{case1}The Delaunay triangulation $T$ contains the point at $\infty$, $v_\infty$ as one of the fixed points. As depicted in fig.~\ref{pointinfty-2}, the faces that do not contain the point at $\infty$ are contained within the convex hull of the Delaunay triangulation of the remaining points $z_v\neq\infty$. The faces that contain  $v_\infty$ are of the form $f=(v,v',v_\infty)$ with $e=(v,v')$ an edge of the convex hull,, and hence are such that $\mathrm{Vol}(f)=0$. They do not contribute to $\mathcal{A}_T$ which becomes
\begin{equation}
\label{ATzinfty}
\mathcal{A}_T\ =\ -\sum_{\begin{smallmatrix}\text{triangles}\ f \\   \subset\,\text{convex hull}\end{smallmatrix}} \mathrm{Vol}(f)
\end{equation} 
 
  \item[Fixed face:]
   \label{case2} The 3 fixed points $(v_1,v_2,v_3)$ are the vertices of a triangle $f_0$ of $T$. Then, by a well chosen SL(2,$\mathbb{C}$) transformation, this face can be taken to be the exterior face of $T$, namely all the other points are inside the (circumcircle of the) exterior face $f_0$. Then $\mathcal{A}_T$ is 
\begin{equation}
\label{AT3z}
\mathcal{A}_T\ =\ \left(-\sum_{\begin{smallmatrix}\text{triangles}\ f\neq f_0\end{smallmatrix}} \mathrm{Vol}(f)\right)\ +\ \mathrm{Vol}(f_0)
\end{equation}  
The $+$ sign for the contribution of the exterior face $f_0$  comes from the fact that it is now clockwise oriented, instead of anti-clockwise as the internal faces of $T$. Since the 3 points $z_1,z_2,z_3$ are fixed, this Vol($f_0$) will not play any role in what follows anyway.
\item[General case]
\label{case3}
The general case is to consider an arbitrary Delaunay triangulation of the Riemann sphere, mapped by stereographic mapping on the complex plane, and to associate the algebraic hyperbolic volume $\mathrm{Vol}(f)$ to each triangle. It may be positive or negative depending on the orientation of $f$. 

\end{description}

\paragraph{Remark:} The hyperbolic volume functional $\mathcal{A}_T$ is \emph{not} the hyperbolic volume functional  that appears in the convex minimization problem for circle patterns (finding the vertices $z_v$ from the edge angle $\theta_e$). Indeed in this latter problem, it is the volume associated to the triangles $(v,f,f')$ (one vertex, 2 circumcircle centers) which appear \cite{Rivin1994} \cite{BobenkoSpringborn2004}.

\begin{definition}[Hermitian form]
The space $\mathfrak{D}_{N+3}\simeq\mathbb{C}^N$ of Delaunay triangulations with three fixed point $(z_1,z_2,z_3)$ (in one of the first two above discussed cases), parametrized by the $N$ remaining points $\mathbf{z}=(z_4,\cdots, z_{N+3})$ is embodied with the Hermitian form $D$
\begin{equation}
\label{Dkd2zA}
D_{u\bar v}(z)\ =\ {\partial\over\partial z_u}{\partial\over\partial \bar z_v}\mathcal{A}_T(z)
\end{equation}
\end{definition}
This form is defined locally for non-coinciding points and in each sub-domain of $\mathbb{C}^N$ corresponding to a given triangulation $T$.
One has the two following results:
\begin{theorem}[$D$ as a Kähler form]
\label{DasKahler}
The Hermitian form $D$ is positive, and continuous on $\mathfrak{D}_{N+3}$, away from coinciding points configurations. Hence it is a continuous Kähler form, whose prepotential is $\mathcal{A}_T$.
\end{theorem}
The proof of Th.~\ref{DasKahler} is given in Sec.~\ref{sGeometryD}, where the explicit geometrical form of the matrix $D$ is given and its conformal properties are discussed.

\begin{theorem}[The measure as a Kähler volume form]
\label{TDasKha}
The measure over Delaunay triangulation is the volume form of $D$
\begin{equation}
\label{DdetDk}
\mathcal{D}_T(z)_{\setminus\{1,2,3\})}\ =\ {\color{black}{2^N}}\ \det\left[
{\left(D_{u,\bar v}\right)}_{\scriptscriptstyle{{\!\!\begin{smallmatrix}\ \\u,v\, \neq \\\{1,2,3\}\end{smallmatrix}}}}
\right]
\end{equation}
\end{theorem}
This means that the measure  $\mathcal{D}_T(z)_{\setminus\{1,2,3\})}$, which is defined as the determinant of a  $2N\times 2N$ real Jacobian matrix, can be written as a simpler $N\times N$ complex determinant.
The proof of Th.~\ref{TDasKha} is given in Sec.~\ref{proofDasKha}.

\subsection{Geometrical form of $D$ and conformal properties}
\label{sGeometryD}
Elementary plane geometry, using the differential of the hyperbolic volume function \ref{dVolH} leads to the explicit form for the matrix elements $D_{u\bar v}$ of $D$. First $D$ can be decomposed into a sum of contributions for each triangle $f$ of $T$.
\begin{equation}
\label{DsumDf1}
D_{u,\bar v}= \sum_{f} D_{u,\bar v}(f)\quad,\qquad D_{u,\bar v}(f)\ =-\ {\partial^2\over\partial z_u\partial \bar z_v}\mathrm{Vol}(f)
\end{equation}
Each $D_{u,\bar v}(f)$ is non-zero only if the vertices $u$ and $v$ belongs to the triangle $f$.

\begin{figure}[h!]
\begin{center}
\includegraphics[width=3in]{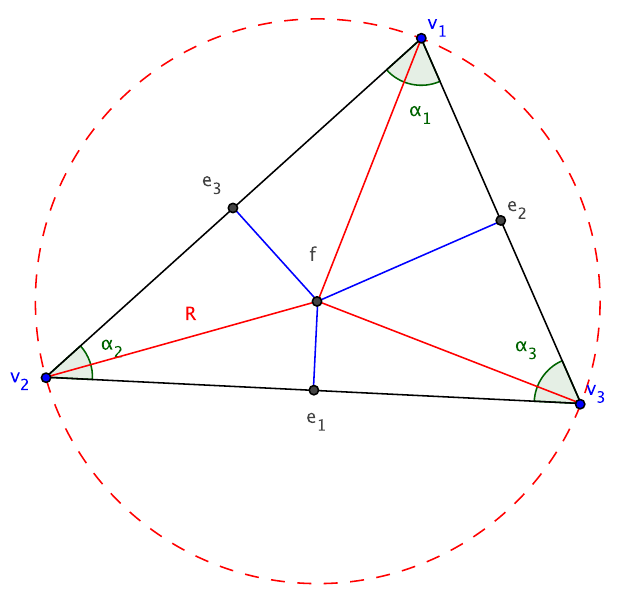}
\caption{A triangle $f=(v_1,v_2,v_3)$. ($e_1,e_2,e_3)$ denote both the edges of $f$ and the middle point of the edges. $f$ denote both the triangle and the center of the circumcircle. $R$ is the radius of the circumcircle.}
\label{triangleS1}
\end{center}
\end{figure}

For a given triangle $f$, depicted on Fig.~\ref{triangleS1},
with (anti clockwise oriented) vertices $(v_1,v_2,v_3)$,  oriented angles $(\alpha_1,\alpha_2,\alpha_3)$ and circumcircle radius $R=R(f)$, the $3\times 3$ matrix $D(f)$ reads
\begin{equation}
\label{DfExpl1}
D(f)\ =\ {1\over 8\,R(f)^2}\begin{pmatrix}
     \cot(\alpha_2)+\cot(\alpha_3) &  -\cot(\alpha_3)-\mathrm{i}  & -\cot(\alpha_2)+\mathrm{i}  \\
     -\cot(\alpha_3)+\mathrm{i}&  \cot(\alpha_3)+\cot(\alpha_1) & -\cot(\alpha_1)-\mathrm{i}   \\
     -\cot(\alpha_2)-\mathrm{i}  & -\cot(\alpha_1)+\mathrm{i} &  \cot(\alpha_1)+\cot(\alpha_2)
\end{pmatrix}
\end{equation}
This can be rewritten as a real and imaginary part
\begin{equation}
\label{DfDelE}
D(f)={- 1\over 8\,R(f)^2}\big( 2\,\Delta_0(f)+ \,\mathrm{i}\, E(f)  \big)
\end{equation}
where $\Delta_0(f)$ is symmetric real,  and $E(f)$ the totally antisymmetric Levi-Civita tensor. $\Delta_0(f)$ is nothing but the contribution of the triangle $f$ to the discretized scalar Laplace-Beltrami operator on the triangulation $T$ in the plane. 

It is easy  to check that $D(f)$ has 2 zero eigenvalues $\lambda_1=\lambda_2=0$, with right eigenvectors $(1,1,1)$ and $(\bar z_1,\bar z_2, \bar z_3)$, and a non trivial positive one (with right eigenvector $(\bar z_1^2,\bar z_2^2, \bar z_3^2)$):
\begin{equation}
\label{positive}
\lambda_3={1\over 4 R^2}(\cot(\alpha_1)+\cot(\alpha_2)+\cot(\alpha_3))\ >0\quad\text{if $f$ non-flat \& counterclockwise oriented}
\end{equation}
From $D(f)\ge 0$ and \ref{DsumDf1}, the positivity of the Hermitean form $D_T$ follows.

\begin{figure}[h!]
\begin{center}
\includegraphics[width=2in]{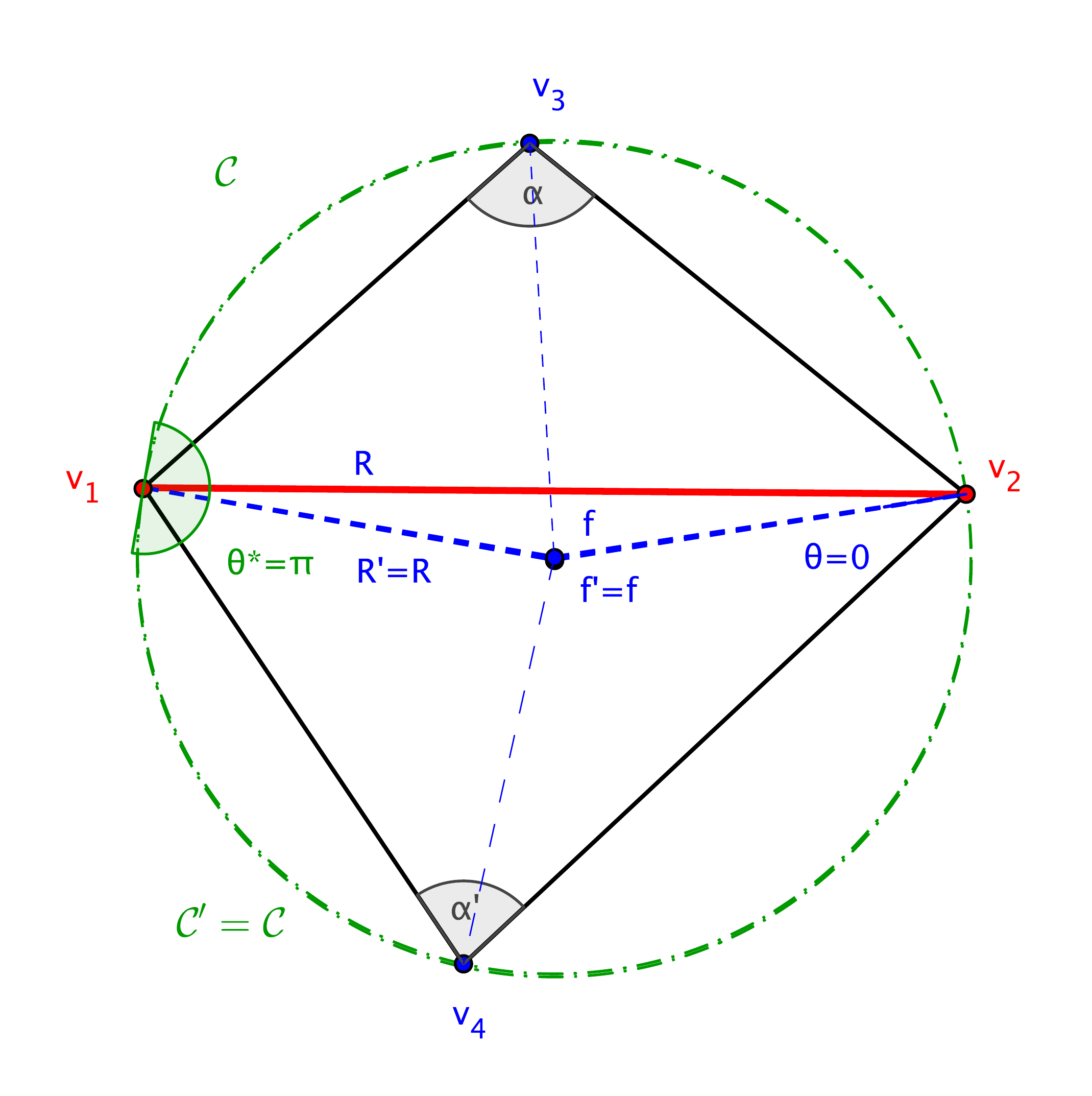}\qquad
\includegraphics[width=2in]{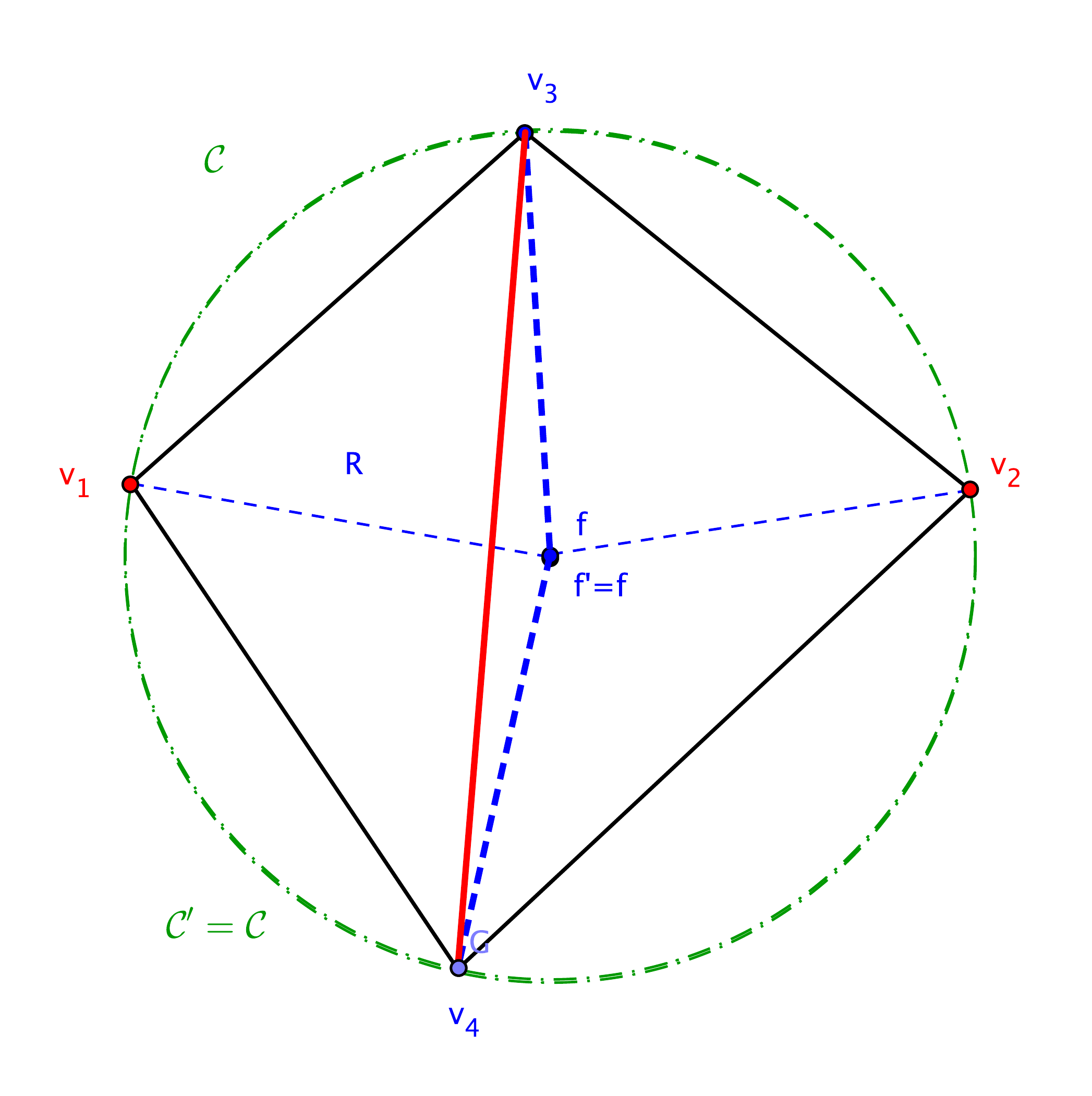}
\caption{A flip}
\label{FlipDet}
\end{center}
\end{figure}

To establish the continuity of $D$, one shows that the matrix elements of $D$ are continuous when a flip of an edge occurs.
As depicted in Fig.~\ref{FlipDet}, the flip of an edge $e=(v_1,v_2)\to e'=(v_3,v_4)$ occurs when the two triangles $f=(v_1,v_2,v_3)$ and $f'=(v_2,v_1,v_4)$ have the same radii $R=R'$, so that their respective centers coincide $w_f=w_{f'}$.
Then it is easy to check from \ref{DfExpl1} that the matrix elements for the edge $e$ vanish, as well as those for the flipped edge $e'$ on the flipped triangulation $T'$,
\begin{equation}
\label{Dnull}
D_{v_1\bar v_2}=0\quad,\qquad D_{v_2\bar v_1}=0
\end{equation}
while the other matrix elements $D_{u\bar v}$ and $D'_{u\bar v}$ are unchanged. This establishes Th.~\ref{DasKahler}

Another simple property of the matrix $D(f)$ for a triangle $f=(z_1,z_2,z_3)$ is
\begin{equation}
\label{zdbzA}
\sum_{i,j=1}^{3} z_i^2\, D_{i\bar\jmath}(f)\, \bar z_j^2\ =\  {1\over 2}\,\mathrm{Area}(f)
\end{equation}
with $\mathrm{Area}(f)$ the algebraic area of the triangle $f$. 

Using Th.~\ref{TDasKha} the measure determinant $\mathcal{D}$ can be shown to have nice properties under global confomal SL(2,$\mathbb{C}$) transformations. For instance, let us define $D_{\setminus a,b,c}$ the $N\times N$ matrix obtained from the general form $D_{u\bar v}$ by removing three (distinct) arbitrary points $(v_a,v_b,v_c)$ of the triangulations $T$. Comparing two different choices of these 3 points (which fix the SL(2,$\mathbb{C}$) invariance), one has
\begin{proposition}
\label{pHdens}
The function $H(z)$ defined as
\begin{equation}
\label{Hdens}
H=
{ \det\left( D_{\backslash_{\scriptscriptstyle{a,b,c}}}(z) \right)   \over \left|\Delta_3(z_a,z_b,z_c)\right|^2}
\end{equation}
with $\Delta_3$ the Vandermonde determinant for the 3 points
\begin{equation}
\label{Vdm3}
\Delta_3(z_a,z_b,z_c)=(z_a-z_b)(z_a-z_c)(z_b-z_c)
\end{equation}
is independent of the choice of the three fixed points $(a,b,c)$. Moreover, it is a weight (1,1) function under global SL(2,$\mathbb{C}$) conformal transformations, 
\begin{equation}
\label{Moebius}
z\to w={az+b\over c z+d}
\quad\text{with}\quad ad-bc=1
\quad,\qquad w'(z)={\partial w\over\partial z}
={1\over (cz+d)^2}
\end{equation}
namely
\begin{equation}
\label{HMtransf}
H(z)
=\left|  \prod_{i=1}^{N+3}{w'(z_i)} \right|^2 H(w)
= \prod_{i=1}^{N+3}   {1\over \left|c z_i+d  \right|^2 } H(w)
\end{equation}

\end{proposition}

\subsection{$\mathcal{D}$ as a discretized Faddeev-Popov operator}
Another nice representation of the Kähler form $D$ leads to the connexion with the continuous formulations of 2 dimensional gravity. Let us define a discretized complex derivative operator on Delaunay triangulations.
\begin{definition}
Let $\mathbb{C}^{\mathcal{V}(T)}$ and $\mathbb{C}^{\mathcal{F}(T)}$ be respectively the vector spaces of complex functions over the vertices and faces (triangles) of a Delaunay triangulation $T$ in the complex plane. We define the complex derivative operators
$\nabla$ and $\overline\nabla$ from $\mathbb{C}^{\mathcal{V}(T)}\to\mathbb{C}^{\mathcal{F}(T)}$ as follows, for an anticlockwise oriented triangle $f=(v_1,v_2,v_3)$
\begin{equation}
\label{nablaZ}
\nabla \Phi(f)\ =\ {1\over 4\ii}{\Phi(v_1)(\bar z_3-\bar z_2)+\Phi(v_2)(\bar z_1-\bar z_3)+\Phi(v_3)(\bar z_2-\bar z_1)\over \mathrm{Area}(f)}
\end{equation}
\begin{equation}
\label{nblaZb}
\overline\nabla \Phi(f)\ =\ -{1\over 4\ii}{\Phi(v_1)(z_3-z_2)+\Phi(v_2)(z_1-z_3)+\Phi(v_3)(z_2-z_1)\over \mathrm{Area}(f)}
\end{equation}
with (again) $\mathrm{Area}(f)$ the algebraic area of the triangle $f$
\begin{equation}
\label{Area2}
\mathrm{Area}(f)= {1\over 4 \ii}\left((z_3-z_1)(\bar z_2-\bar z_1)-(z_2-z_1)(\bar z_3-\bar z_1) \right) 
\end{equation}
\end{definition}
This corresponds to the naive definition of the derivative of the function $\Phi$  pointwise defined on the vertices $z_i$ of the triangulation $T$, and interpolated linearly inside each triangle. Indeed, if $\Phi$ is linear over $\mathbb{C}$, $\Phi(v_i)=a+b\,z_i+ c\,\bar z_i$, then $\nabla\Phi= b$ and $\overline\nabla\Phi=c$.
These operators $\nabla$ and $\overline\nabla$ differ from the $\partial$ and $\bar\partial$ operators usually considered in the theory of discrete analytic functions (as far as we know, it is not possible to define standard discrete analyticity on generic Delaunay triangulations).

The scalar product of the local operator $D(f)$ relative to a single triangle $f$ between two functions of $\mathbb{C}^{\mathcal{V}(T)}$  takes the specific form
\begin{proposition}
\label{pPhiDPsi}
\begin{equation}
\label{PhiDPsi}
\Phi \cdot D(f)  \cdot\overline\Psi
\ =\ \sum_{i,j\,\mathrm{vertices\, of}\, f}\Phi(v_i) D_{i\overline\jmath}(f)\overline\Psi(v_j)
\ =\ {\mathrm{Area}(f)\over R(f)^2}\  \overline{\nabla} \Phi(f)\  \nabla\overline\Psi(f)
\end{equation}
\end{proposition}
This formula is very suggestive. Summing over the triangles we can rewrite it (in short hand notations) as
\begin{equation}
\label{DnabAR2nab}
D=\ \sum_fD(f)\ =\ \nabla^\dagger {A\over R^2} \nabla
\end{equation}
Remember that $D=\sum_fD(f)$ is a complex Kähler form and can be viewed as a linear operator acting on the space of real vector fields $V$ (living on the vertices of $T$). For this, we identify a complex function $\Psi\in\mathbb{C}^{\mathcal{V}(T)}$ with a real vector field  with  components in the coordinate $z=x^1+\ii \,x^2$
\begin{equation}
\label{Psixy}
(\Psi^1,\Psi^2)=(\Re(\Psi),\Im(\Psi))
\end{equation}
or in complex component notations
\begin{equation}
\label{Psizzbar}
(\Psi^z,\Psi^{\bar z})=(\Psi,\overline\Psi)
\end{equation}
Now let us denote  $w_f$ the complex coordinate of the center of the circumcircle of the triangle $f$, the area of the triangle  as a volume element
\begin{equation}
\label{AreaW}
\mathrm{Area}(f)\ =\ d^2 w_f
\end{equation}
and the $R(f)^{-2}$ factor as a ``quantum area'' Liouville factor
\begin{equation}
\label{LiouvPhi}
{1\over R(f)^2}=\e^{\phi(w_f)}
\end{equation}
Then we can rewrite formally (by summing \ref{PhiDPsi} over the faces) the scalar Kähler product of two vector fields as an integral
\begin{equation}
\label{DasFP}
\Phi\cdot D\cdot\overline\Psi\ =\ \int d^2w\ \e^{\phi(w)}\ \partial_{\bar z}\Phi^z(w)\ \partial_z\Psi^{\bar z}(w)
\end{equation}

This is very reminiscent of the conformal gauge fixing Fadeev-Popov operator introduced by Polyakov in his famous 1981 paper
\cite{Polyakov1981207}, see \cite{Friedan:1982is}  for details.
Remember that the functional integral over 2-dimensional Riemanian metrics $g_{ab}(z)$ is performed by choosing a conformal gauge
\begin{equation}
\label{ConfG}
g_{ab}(z)=\delta_{ab}\ \e^{\phi(z)}
\end{equation}
This gauge fixing introduces a Faddev-Popov determinant in the functional measure 
\begin{equation}
\label{GMeas}
\int\mathcal{D}[g_{ab}]\ =\ \int\mathcal{D}[\phi]\ \det(\nabla_{\scriptscriptstyle\mathrm{\!FP}})
\end{equation}
where $\nabla_{\scriptscriptstyle\mathrm{\!FP}}$ is the differential operator that maps vector fields $C=\{C^a\}$ onto traceless symmetric tensors $B=\{B^{ab}\}$ via ($D^a$ being the covariant derivative)
\begin{equation}
\label{FPD}
B^{ab}=(\nabla_{\scriptscriptstyle\mathrm{\!FP}}C)^{ab}= D^a C^b + D^b C^a -g^{ab} D_c C^c
\end{equation}
In the conformal gauge and in complex coordinates one has $C=(C^z,C^{\bar z})$ and
\begin{equation}
\label{FPDz}
(\nabla_{\scriptscriptstyle\mathrm{\!FP}}C)^{zz}=\e^{-\phi}\  \partial_{\bar z} C^z
\ ,\quad
(\nabla_{\scriptscriptstyle\mathrm{\!FP}}C)^{\bar z\bar z}=\e^{-\phi}\  \partial_{z} C^{\bar z}
\ ,\quad
(\nabla_{\scriptscriptstyle\mathrm{\!FP}}C)^{z\bar z}=
(\nabla_{\scriptscriptstyle\mathrm{\!FP}}C)^{\bar z z}=0
\end{equation}
The determinant is usually computed by introducing a ghost-antighost system
\begin{equation}
\label{bcsystem}
\mathbf{c}=(c^z,c^{\bar z})=(c,\bar c)\ ,\quad \mathbf{b}=(b_{zz},b_{\bar z\bar z})=(b,\bar b)
\end{equation}
and writing
\begin{equation}
\label{ FPdetbc}
\det(\nabla_{\scriptscriptstyle\mathrm{\!FP}})\ =\ \int \mathcal{D}[\mathbf{c},\mathbf{b}]\ \exp\left({\int d^2 z\, \e^{\phi}\, \left(b_{zz} (\nabla c)^{zz} + b_{\bar z\bar z}  (\nabla c)^{\bar z\bar z}\right)} \right)
\end{equation}
In the standard approach one treats separately the holomorphic $(b,c)$ and anti-holomorphic $(\bar b,\bar c)$ ghost fields. Each of them is a conformal field theory and contribute by a central charge $c=-13$ to the conformal anomaly.
One may however integrate over the anti-ghosts $\mathbf{b}=(b,\bar b)$, keeping only the ghosts $\mathbf{c}=(c,\bar c)$. One then obtains
\begin{equation}
\label{FPdetc}
\det(\nabla_{\scriptscriptstyle\mathrm{\!FP}})\ =\ \int \mathcal{D}[\mathbf{c}]\ \exp\left({\int d^2 z\, \e^{\phi}\, \partial_z c^{\bar z}\ \partial_{\bar z}c^z} \right)
\end{equation}
One recognizes the Kähler form $c{\cdot}D{\cdot}\bar c$ that appears in \ref{DasFP}. Therefore the Kähler operator $D$ defined on the Delaunay triangulation $T$ is in fact a correct discretization of the conformal gauge Fadeev-Popov operator of 2 dimensional gravity
\begin{equation}
\label{DisFP}
D\ =\ \nabla_{\scriptscriptstyle\mathrm{\!FP}}
\end{equation}
while the field $\phi(f)$ defined by \ref{LiouvPhi} in term of the radii of the triangles
\begin{equation}
\label{PhiLogR}
\phi(f)=-\,2\,\log(R(f))
\end{equation}
and which lives on the vertices of the Voronoï lattice, dual to the Delaunay triangulation $T$, is an attractive discretization of the Liouville field of the continuous formulation.

\subsection{Relation with planar maps and $c=0$ 2d gravity}
\subsubsection{Intrinsic planar triangulations and local curvature}
\begin{figure}[h]
\begin{center}
\includegraphics[width=7.in]{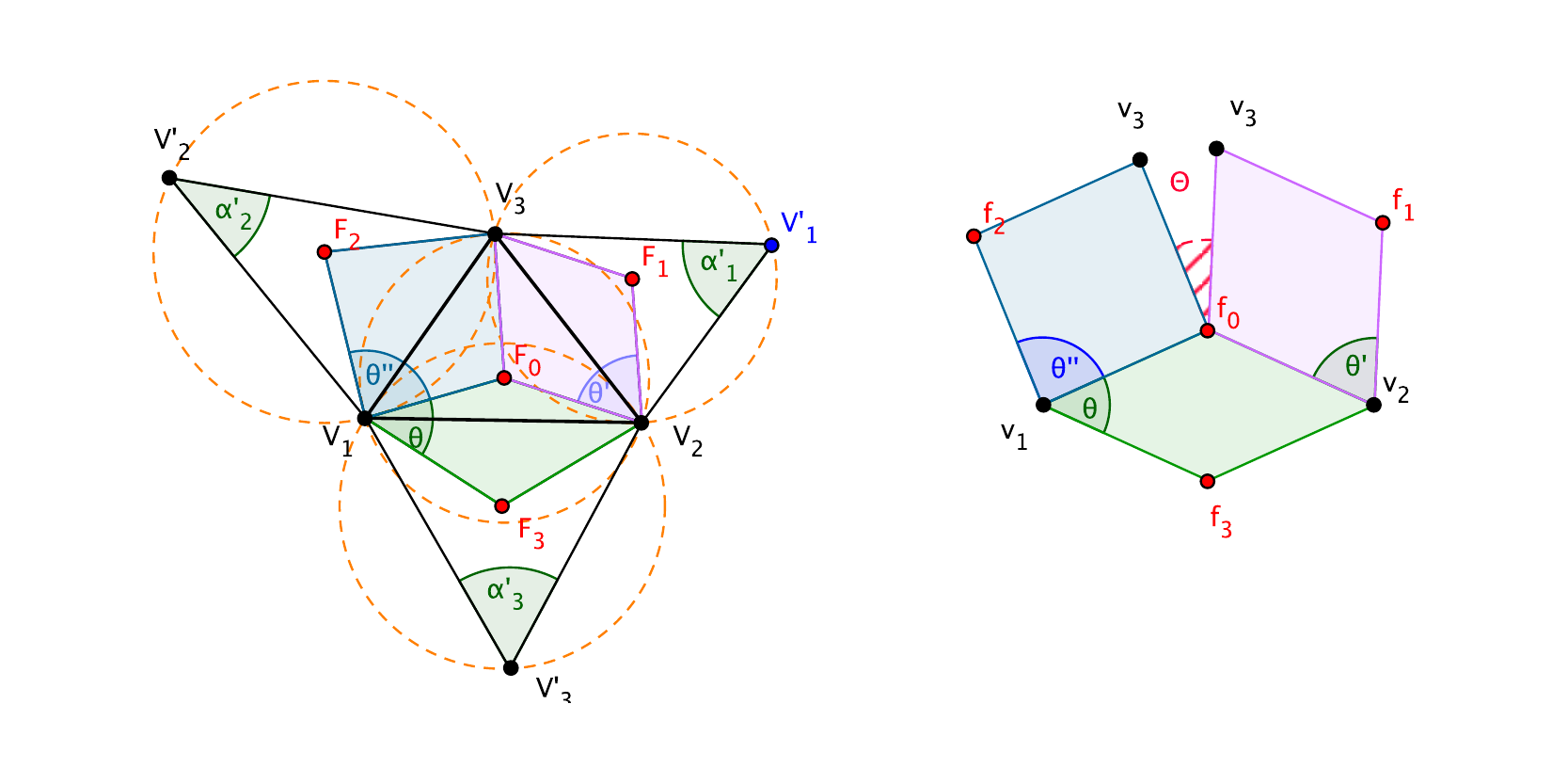}
\caption{Kite decomposition of a Delaunay triangulation (left), and the the corresponding rhombi complex (right) with unit edge length ($\ell=1$) and curvature at the $f$ vertices corresponding to the center of the faces (triangles).}
\label{4triangles}
\end{center}
\end{figure}

Let us discuss the relation between our model and other models of abstract planar maps or random triangulations.
Consider a given triangle $F_0$ of a Delaunay triangulation $\tilde T=(T,\theta)$, as depicted on Fig.~\ref{4triangles} (left), with its 3 neighbors $F_1$, $F_2$ and $F_3$. Consider the kite parallelogram $K=(V_1,F_3,V_2,F_0)$ associated to the edge $e=(V_1,V_2)$. It can be mapped (by a SL(2,$\mathbb{C}$) transformation) onto a rhombus 
$R=(v_1,f_3,v_2,f_0)$ with unit edge length $\ell=|v_1f_0|=1$ with the same edge angle $\theta=\widehat{(F_3 V_1 F_0)}=\widehat{(f_3 v_1   f_0)}$. Repeating this operation for all the edges, we obtain a planar, but non-flat complex of rhombi, with angles $\theta$ (at the $v$ vertices) and $\theta^*=\pi-\theta$ (at the $f$ vertices). Condition \ref{sumthetav} implies flatness at the $v$ vertices, but there is a defect angle $\Theta_f$ at the $f$ vertices, given for the vertex $f_0$ by 
\begin{equation}
\label{Thetaf}
\Theta_f=\theta+\theta'+\theta''-\pi=\pi-\alpha'_1-\alpha'_2-\alpha'_3
\end{equation}
that we can identify with local scalar curvature $\Theta_f$ localized at the center $w_f$ of the circumcircle $\mathcal{C}_f$ of the triangle $f$ in the Delaunay triangulation.

So our model can be formulated as a sum over planar quadrangulations (hence a bipartite lattice with $f$ and $v$ vertices), with a Euclidean structure (given by the angles of the rhombi), with the two constraints : (1) flatness, i.e. $\sum\theta=2\pi$, at the $v$ vertices, (2) only 3 rhombi meet at $f$ vertices.

\begin{figure}
\begin{center}
\includegraphics[width=3in]{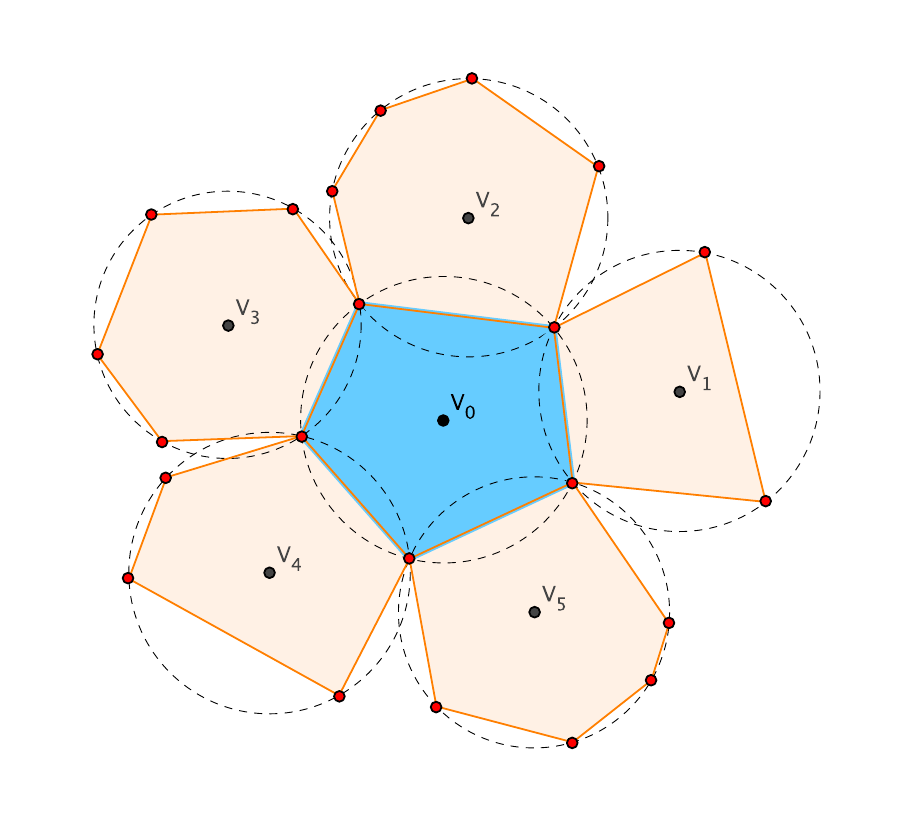}
\caption{Equivalent decomposition into planar isoradial polygons}
\label{6polygons}
\end{center}
\end{figure}

One may instead decompose the complex into polygons centered around  the $v$ vertices, whose summits are the neighbors $f$ vertices, as depicted in Fig.~\ref{6polygons}. The model is then formulated as a sum over planar polyhedra complex with the three constraints: (1) all polyhedra are isoradial, i.e. their vertex belongs to a circumcircle with unit radius $\ell=1$, (2) they are glued along their edges, (3) only three polyhedra meet at a vertex.

\subsubsection{Relation with other models}
The main point is that the model is a variation of the model of planar triangulations (or the dual model of planar trivalent maps), with some geometrical constraint on the vertices and the faces. It is known that unless some very specific constraints are applied, these models belong to the same universality class than the simple random planar map model), namely the class of $c=0$ pure two dimensional gravity. So we conjecture that our model also belong to this class. Of course a full solution is needed to confirm this claim, but this is indeed the case in some limiting case.

Let us for instance modify the model by changing the flat measure over the $\theta_e$'s onto a measure of the form
\begin{equation}
\label{ }
\prod_e d{\theta_e}\  |\cos(\theta_e)|^{2k}\ \prod_v \delta\left(\sum_{e\to v} \theta_e -2\pi\right)
\end{equation}
In the limit $k\to\infty$, the only triangulations $(T,\boldsymbol{\theta})$ that survive are those such that
\begin{equation}
\label{ }
\theta_e=0\quad\text{or}\quad\pi
\end{equation}
and the constraint $\sum\limits_{e\to v} \theta_e =2\pi$ implies that there are 2 and only 2 edges such that $\theta_e=\pi$ at each vertex. In this limit, the model reduces to a combinatorial model of fully packed loop systems on random triangulations (of course the Euclidean metric on such a triangulation becomes degenerate). No particular weight is attached to the loop system, since each loops configuration is counted with true same weight. Therefore the model corresponds to a fully packed loop model with $z=1$ loop fugacity, i.e. to a dense O($N$) model for $N=1/2$.
The loops live on the edges $e$  of the triangulation $T$, not on the edges $e^*$ of the dual trivalent lattice $T^*$. $T$ is generically not an Eulerian lattice, and there should be no difference between the fully packed model and the densely packed model. Densely packed loop models  with $z=1$ loop fugacity are known to be described in the continuum limit (i.e. at large distances) by a $c=0$ conformal field theory. This was our claim.

\subsubsection{Isoradial limit}
Another simple limit is to enforce the constraint that the local curvature $\Theta_f$ given by \ref{Thetaf} vanishes for all faces (not discussing the boundary conditions). Then it is easy to see that this is equivalent to the isoradiality condition that (in a proper coordinate system) all the circumcircle radii $R_f$ are equal
\begin{equation}
\label{isorad}
\Theta_f=0\quad\forall\ f\qquad \iff\qquad R_f=\ell  \quad\forall\ f\qquad
\end{equation}
In this limit, only flat rhombic planar graph survives, and from \ref{PhiDPsi} the discretized Fadeev-Popov operator $D$ reduces to the scalar Laplace-Beltrami operator $\Delta_0=4\bar\partial\partial$
\begin{equation}
\label{isoradlim}
\text{isoradial limit}\quad\implies \qquad D = - {1\over 4\,\ell^2}\ \Delta_0
\end{equation}
This case has been much studied in connection with dimer models and discrete analyticity, and there exists beautiful explicit forms for the determinant  of $\Delta_0$ and its inverse (the scalar propagator) \cite{Kenyon:2002uq,KenyonSchlenker2004} .

\section{Link with topological gravity}

Another approach to quantum gravity, is "topological gravity", introduced by Witten \cite{Witten:1990}, and culminating with Kontsevich's proof  \cite{Kontsevich1992} .
Here we make the link between our approach and Witten-Kontsevich, by relating our measure ${\mathcal{D}_T(z)}_{\setminus\{1,2,3\}}$ to a combination of Chern classes on the moduli space of Riemann surfaces.

\subsection{Chern classes}

${\mathfrak{T}}_{N+3}=\mathbb C^{N+3}/Sl_2(\mathbb C)$ the set of $N+3$ point on the Riemann sphere, is the moduli space ${\cal M}_{0,N+3}$ of Riemann surfaces of genus zero with $N+3$ marked points, and we have seen that it is isomorphic to the combinatorial space $\widetilde{\mathfrak{T}}_{N+3}$ of abstract triangulations $T$ with angles $\theta_e$ associated to edges with the constraint that $\sum_{e\mapsto v} \theta_e = 2\pi$.
A good tool to study the topology of such a space, is to consider Chern classes of $U(1)$ bundles.

Let us define the following circle bundle: ${\cal L}_v \to {\mathfrak{T}}_{N+3}$, as follows:
the fiber over the point $\{z_1,\dots,z_{N+3}\}\in\widetilde{\mathfrak{T}}_{N+3}$, is the unit circle $S_v$ of center $v$ in the Euclidian plane. Some special points on $S_v$ are the intersections $p_{v,f}=S_v\cap [v,f)$ of $S_v$ with the half lines $[v,f)$ emanating from $v$ towards the centers $f$ of adjacent circumcircles.
\begin{figure}[h]
\begin{center}
\includegraphics[width=4in]{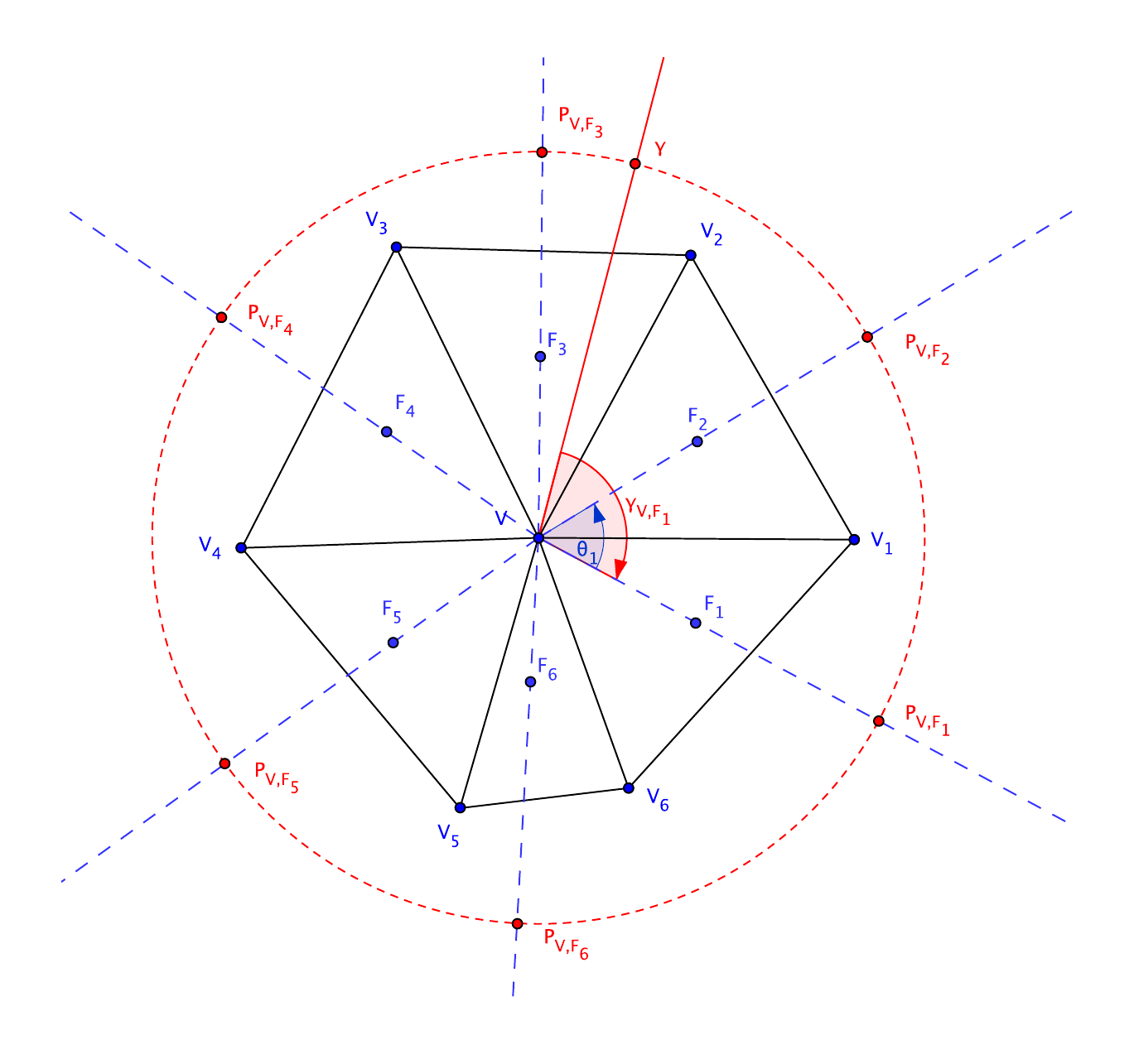}
\caption{The fiber of the circle bundle ${\cal L}_v\to {\mathfrak{T}}_{N+3}$, is a circle centered at $v$. A point in the fiber, is a point $\gamma$ on the circle, and a coordinate for that point is the angle $\gamma_{f,v}$ between $\gamma$ and the segment $[v,f]$ where $f$ is a face adjacent to $v$.}
\label{Chern}
\end{center}
\end{figure}
Consider a section $\gamma\in S_v$. For each face $f$ adjacent to $v$,  define the angles:
\beq
\gamma_{v,f} = {\rm angle\,along}\,S_v\,\,{\rm between}\,\,\gamma\,{\rm and}\,p_{v,f}
\eeq
If we label the faces $f_1,\dots,f_n$ in the trigonometric order arounf $v$, we have:
\beq
\gamma_{v,f+1} = \gamma_{v,f}-\theta_{f_+} = \gamma_{v,1} - \theta_{1_+} - \dots - \theta_{f_+}
\eeq
where $[v,f_+]$ is the edge  following $[v,f)$ around $S_v$.
The following 1-form depends only on $v$, it is independent of the choice of labelling of faces around $v$:
\beq
u_v=\frac{1}{4\pi^2} \sum_{f\mapsto v}\, \theta_{v,f_+} d\gamma_{v,f}
= \frac{d\gamma_{v,1}}{2\pi} - \frac{1}{4\pi^2} \sum_{f=2}^n\sum_{f'=1}^{f-1} \theta_{v,f_+} d\theta_{v,f'_+}
\eeq
It is clear that 
\beq
\int_{S_v} u_v
=\frac{1}{4\pi^2} \sum_{f\mapsto v}\, \theta_{v,f_+} \int_{S_v} d\gamma_{v,f}
=\frac{2\pi}{4\pi^2} \sum_{f\mapsto v}\, \theta_{v,f_+} 
= 1
\eeq
In other words, $u_v$ is a connection globally defined on the bundle ${\cal L}_v$, whose integral along the fiber is $1$, its curvature $du_v$ is then the Chern class:
\beq
\psi_v = c_1({\cal L}_v) = du_v =  - \frac{1}{4\pi^2} \sum_{f=2}^n\sum_{f'=1}^{f-1} d\theta_{v,f_+} \wedge d\theta_{v,f'_+}
\eeq
The Chern class $\psi_v$ of the bundle ${\cal L}_v$ is a 2-form on ${\mathfrak{T}}_{N+3}$, notice that it is independent of a choice of origin of labelling of faces around $v$.

In other words, if we label the edges around $v$ in the trigonometric order $e_1,\dots,e_n$, we have:
\beq
\psi_v = - \frac{1}{4\pi^2} \sum_{e=2}^n\sum_{e'=1}^{e-1} d\theta_{e} \wedge d\theta_{e'}
\eeq
Notice that $\psi_v$ is independent of the choice of labelling, and also, notice that since $\sum_{e\mapsto v} \theta_e=2\pi$, we also have:
\beq
\psi_v = - \frac{1}{4\pi^2} \sum_{e=2}^{n-1}\sum_{e'=1}^{e-1} d\theta_{e} \wedge d\theta_{e'}
\eeq
where the upper bound of the sum is now $e\leq n-1$, instead of $n$.

\subsection{Measure and Chern classes}

The $2N$ form $(\sum_v \psi_v)^N$ is a top-dimensional form on ${\mathfrak{T}}_{N+3}$, with constant coefficients, therefore it must be proportional to $\prod_{e\in {\cal E}_0} d\theta_e$, i.e. to our measure ${\mathcal{D}_T(z)}_{\setminus\{1,2,3\}}$:
\beq
(\sum_{v=1}^{N+3} \psi_v)^N = C_T\,\,{\mathcal{D}_T(z)}_{\setminus\{1,2,3\}}
\eeq
Notice that the coefficient $C_T$ might apparently depend on the triangulation $T$, however, it was proved by Kontsevich that it doesn't  %
\footnote{Kontsevich introduced the combinatorial space of Strebel graphs ${\cal M}_{g,n}^{\rm comb}=\oplus_T \mathbb R_+^{\#{\rm edges}(T)}$ with trivalent graphs of genus $g$ and $n$ vertices, with some coordinates $l_e\in\mathbb R_+$ on each edge, subject to constraints $\sum_{e\mapsto v} l_e=L_v$ fixed at each vertex. He considered the Chern classes $\psi_v=\sum_{e'<e\,{\rm around}\,v} dl_e\wedge dl_{e'}$, and proved that $(\sum_v \psi_v)^{3g-3+n}\,\prod_{v} dL_v = (3g-3+n)!\,2^{5g-5+2n}\,\prod_e dl_e$.
Our space ${\cal T}_{N+3}={\cal M}_{0,N+3}^{\rm comb}$ corresponds to the planar case $g=0$ and $n=N+3$, and we identify $l_e=\theta_e/2\pi$. }  %
, and we have:
\beq
\left(\sum_v\,\,4\pi^2\,\psi_v \right)^N = \pm \,\,N!\, 2^{2N+1}\,\,\prod_e d\theta_e
\eeq

This shows that our measure ${\mathcal{D}_T(z)}_{\setminus\{1,2,3\}}$ is also the measure of topological gravity.

\section{Proofs of the results}
\subsection{Chosing a basis of edges}
\subsubsection{Proof of Th.~\ref{ThE0def}}
We have defined the measure on $\td{\cal T}_{N+3}$ to be the uniform measure on triangulations tensored with the flat Lebesgue measure on angles $\theta_e$'s (constrained by \eqref{sumthetav}):
\beq
{\rm uniform}(T)\otimes \prod_{e\in {\cal E}(T)}\,d\theta_e\quad \prod_{v\in {\cal V}(T)}\,\delta(-2\pi+\sum_{e\mapsto v} \theta_e)
\eeq
One takes care of the constraints $\sum_{e\mapsto v} \theta_e =2\pi$ by chosing a basis of $2N$ independent edges.
We now prove Th.~\ref{ThE0def}, which characterizes these basis.

\begin{definition}[Adjacency matrix $R$]
\label{defR}
Let us define the $(N+3)\times (3N+3)$ matrix $R$ by
\beq
\label{defR}
R_{v,e} = \left\{\begin{array}{l}
1\quad {\rm if}\,v\,{\rm adjacent\,to}\, e \cr
0\quad {\rm otherwise}
\end{array}\right.
\eeq
\end{definition}
If we choose a set $\calE_0\subset {\cal E}(T)$ of $2N$ independent edges, we have:
\beq\label{eqmeasuretdTdthetaE0a}
\prod_{e\in {\cal E}(T)}\,d\theta_e\quad \prod_{v\in {\cal V}(T)}\,\delta(-2\pi+\sum_{e\mapsto v} \theta_e)
= \frac{1}{\det_{v\in{\cal V}(T),\, e\in {\cal E}(T)\setminus \calE_0} R_{v,e}} \quad \prod_{e\in \calE_0}\,d\theta_e
\eeq
Assume that we can find a basis $\calE_0$ with $2N$ edges.

First, notice that for every vertex $v$ of $T$, at least one of adjacent edges to $v$ must not be in $\calE_0$ (otherwise the determinant at the denominator in 
\eqref{eqmeasuretdTdthetaE0a} 
would have a vanishing line, and be vanishing).

The dual $\calE_0^*$ of $\calE_0$ (i.e. the set of edges on the dual graph of $T$ crossing the edges of $\calE_0$), must thus contain no loop, i.e. it must be a tree or a union of $k$ disjoint trees on the dual graph.

Since a tree contains one more vertex than edges, and since $\calE_0^*$ contains $2N$ edges, then $\calE_0^*$ must contain $2N+k$ dual vertices i.e. faces (vertices of the dual $\calE_0^*$ are faces  of $T$), but since there are $2N+2$ faces in $T$, we must have $k\leq 2$, i.e. $k=1$ or $k=2$. Notice that if $k=1$, that means that exactly one face of $T$ is not reached by ${\cal E}_0^*$, and one can say that this face, is a tree with 1 vertex and no edge, in some sense we are back to $k=2$ by allowing one of the 2 trees to be reduced to a point.
This shows that $\calE_0^*$ must have two connected components.

\begin{figure}[h!]
\begin{center}
\includegraphics[width=7.cm]{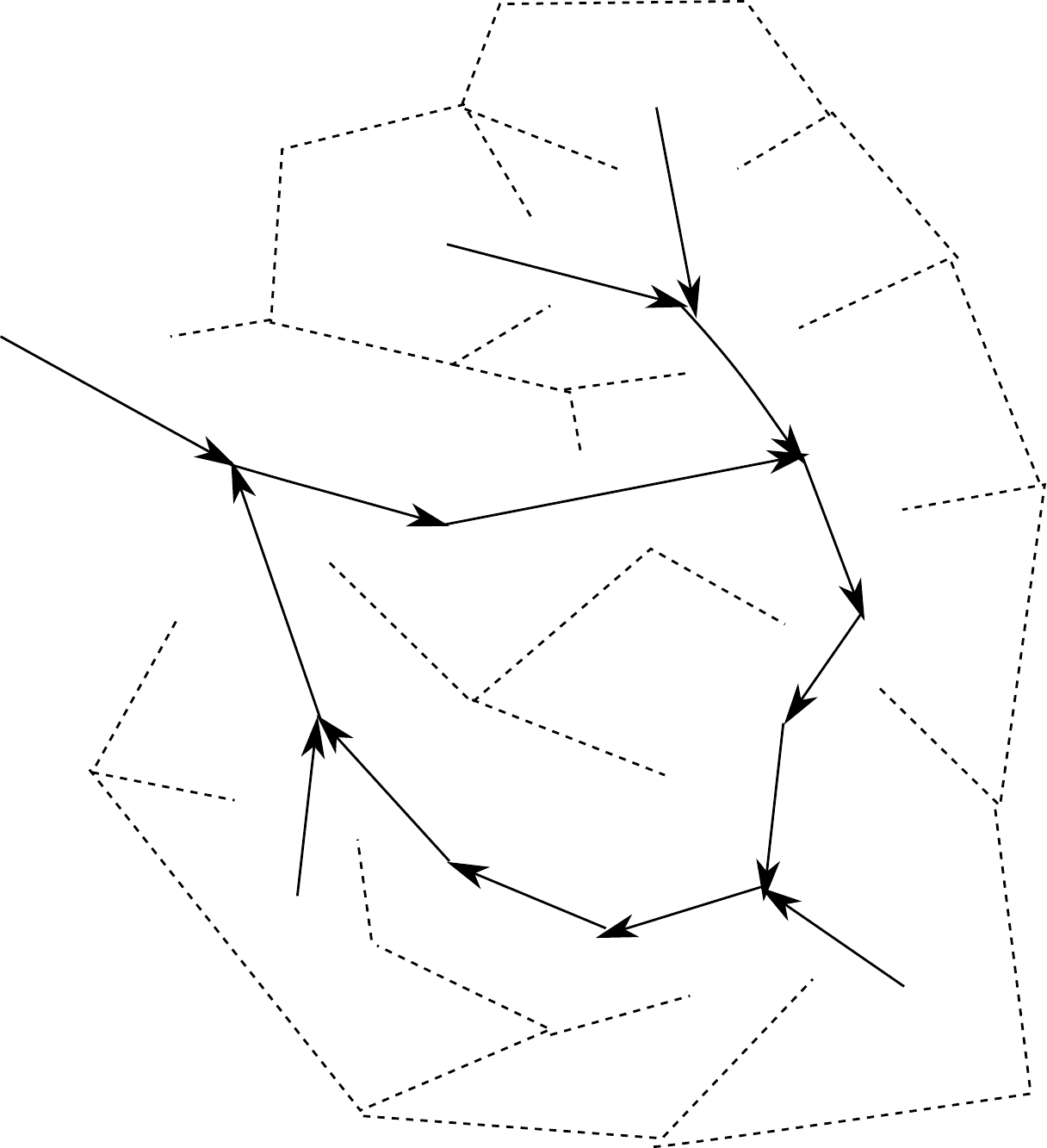}
\caption{A basis ${\cal E}_0$ of $2N$ edges, is such that the complementary $\bar{\cal E}_0={\cal E}(T)\setminus {\cal E}_0$ is a 1-loop rooted tree, whose loop has odd length. Its dual ${\cal E}_0^*$ is made of 2 trees with a total of $2N$ edges.}
\label{figtree1loop}
\end{center}
\end{figure}

This implies that $\bar \calE_0 = {\cal E}(T)\setminus \calE_0$ is a set of $N+3$ edges and must contain exactly one loop.
This implies that $\bar \calE_0$ is a 1-loop rooted tree, i.e. a loop with trees attached to it. See fig.\ref{figtree1loop}.

We now have:
\beq
\label{detRperm}
\det_{v\in {\cal V}(T),\,e\in \bar \calE_0}\,\,R_{v,e}
=\sum_{\sigma: {\cal V}(T)\to \bar \calE_0} (-1)^\sigma\,\,\prod_{v} R_{v,\sigma(v)}
\eeq
A term $R_{v,\sigma(v)}$ is non--vanishing (and is then $=1$) only if $\sigma(v)$ is an edge of $\bar \calE_0$ adjacent to $v$, we represent it as an arrow on edge $e$ with origin $v$.
The determinant is thus a sum over all arrowed paths on $\bar \calE_0$.
There exist only two possible arrowed paths on $\bar \calE_0$, obtained from putting arrows from the leaves of $\bar \calE_0$ towards the loop, and putting arrows around on the loop, in the two possible directions around the loop.
If the loop has  length $l$, reversing the arrows multiplies the signature of $\sigma$ by $(-1)^{l-1}$.
We thus have:
\beq
\label{detRpm2}
\det_{v\in {\cal V}(E),\,e\in \bar \calE_0}\,\,R_{v,e}
= \pm \,\,(1+(-1)^{l-1})
\eeq
i.e. it vanishes if the loop of $\bar \calE_0$ has even length, and is equal to $\pm 2$ otherwise.

A necessary condition for ${\cal E}_0$ to be a basis is thus that $\bar \calE_0 = {\cal E}(T)\setminus \calE_0$ is a 1-loop tree, with a loop of odd length.
Its is also a sufficient condition because then $\det_{v\in {\cal V}(E),\,e\in \bar \calE_0}\,\,R_{v,e} = \pm 2$ is non vanishing.
This ends the proof of Th.~\ref{ThE0def}.

\subsubsection{Example and proof of Th.~\ref{theorem2}}

\begin{figure}[h!]
\begin{center}
\includegraphics[width=12.cm]{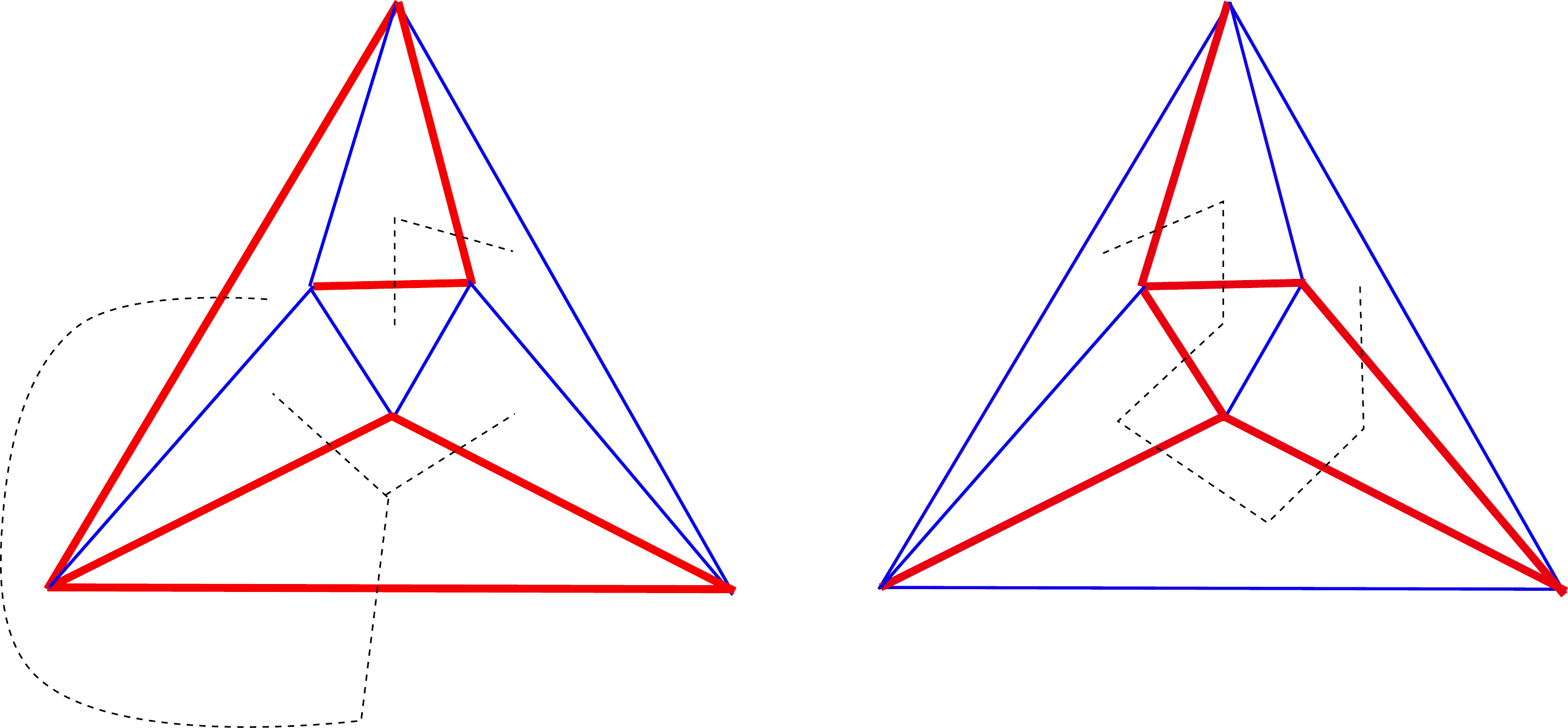}
\caption{2 examples of basis of $6$ edges for the octaedron ($N=3$). In both cases, the basis is made of the $2N=6$ red thick edges, its complementary is a 1-loop rooted tree ($N+3=6$ thin blue edges), and the dual is made of 2 trees ($2N=6$ dashed edges). In the second example, the 1-loop rooted tree is chosen such that the loop is the triangle whose face is the point at $\infty$, and thus one of the 2 dual trees is reduced to the point at $\infty$.}
\label{figbasisocta}
\end{center}
\end{figure}

A particular example is obtained by chosing the loop of ${\bar{\cal E}_0}$ to be a triangle around a face $f_0$ of $T$, i.e. one of the two connected components of ${\cal E}_0^*$ to be a point, namely the center of face $f_0$.
See fig.\ref{figbasisocta}.
In that case, $\calE_0$ is the dual of a covering tree $\calE_0^*$ of the dual graph of $T\setminus f_0$.

Finally, from \ref{eqmeasuretdTdthetaE0a} and  \ref{detRpm2} one has that if $\mathcal{E}_0$ is an admissible basis, 
the measure on $\td{\cal T}_{N+3}$ is
\beq\label{eqmeasuretdTdthetaE0bis}
\frac{1}{2} \,\,{\rm uniform}(T) \,\otimes\,\, \prod_{e\in \calE_0}\,d\theta_e
\eeq
and is 
independent of a choice of basis $\calE_0\subset \calE(T)$.
This is Th.~\ref{theorem2}

\subsection{Computation of the determinants $\det(J)$ and $\det(D)$}
We now show how to compute the Jacobian determinant $J$ and prove the main theorems Th.~\ref{ThDas3T} and Th.~\ref{TDasKha}. We first introduce some important matrices. We chose a given Delaunay triangulation $T$.

\begin{definition}[Edge-edge $E$ matrix]
Let us define the $(3N+3)\times (3N+3)$ matrix $E=\{E_{e,e'}\}$ for edges $(e,e')\in {\cal E}(T)^2$:
\beq
\label{defE}
E_{e,e'} = \left\{
\begin{array}{l}
0 \,\,\,\, {\rm if}\,\,e,e'\,\,{\rm not\,in\,same\,triangle} \cr
+1\,\,\, {\rm if}\,\,e\mapsto e'\,\,\,{\rm clockwise} \cr
-1 \,\,\, {\rm if}\,\,e\mapsto e'\,\,\,{\rm counter-clockwise} 
\end{array}
\right.
\eeq
\end{definition}

\begin{definition}[Edge-vertex $A$ and $\bar A$ matrices]
The $(N+3)\times (3N+3)$ matrix $A=\{A_{v,e}\}$ where $v$ is a vertex of $T$ and $e$ an edge of $T$ is defined as
\beq
\label{defA}
A_{v,e} = 
\begin{cases}
   {1\over z_v-z_{v'}}   & \text{if $e$ is adjacent to $v$, i.e. $R_{v,e}=1$}, \\
    0  & \text{otherwise}.
\end{cases}
\eeq
where $v'$ denotes the vertex of $T$ opposite to $v$ on the edge $e=(v,v')$. The matrix $\bar A$ is the complex conjugate of $A$
\beq
\label{defbarA}
\bar A_{v,e} = 
\begin{cases}
   {1\over \bar z_v-\bar z_{v'}}   & \text{if $e$ is adjacent to $v$, i.e. $R_{v,e}=1$}, \\
    0  & \text{otherwise}.
\end{cases}
\eeq
The matrices $A$ and $\bar A$ appear in the tree expansion of $\det J$ in \ref{Das3trees}.

\end{definition}
We now choose an admissible basis of $2 N$ edges $\mathcal{E}_0\subset\mathcal{E}(T)$.
\begin{definition}[$P_0$ and $M_0$ matrices attached to an edge basis $\mathcal{E}_0$ ]
Let  $P_{0}$ be the $(3N+3)\times (2N)$ matrix projector on ${\cal E}_0$, i.e. (in a basis where $\mathcal{E}_0=(e_1,\cdots,e_{2N})$ and $\bar{\mathcal{E}}_0=\mathcal{E}(T)\setminus\mathcal{E}_0=(e_{2N+1},\cdots e_{3N+3})$)
\begin{equation}
\label{P0def}
P_0=\begin{pmatrix}
   \mathrm{Id}_{2N} \\
      0 
\end{pmatrix}
\end{equation}
Since $\mathcal{E}_0$ is an admissible basis, the $(N+3)\times (3N+3)$ adjacency matrix $R$ defined by \ref{defR} can be decomposed into a $(N+3)\times (2N)$ bloc $R_0$ and a $(N+3)\times(N+3)$ invertible bloc $\tilde{R}_0$
\begin{equation}
\label{RR0Rb0}
R = (\overbrace{R_0}^{\calE_0=2N} |\, \overbrace{{\tilde R}_0}^{N+3})\ \}\scriptstyle{N+3}
\end{equation}
Let us define the matrix $M_0$ as the $(3N+3)\times (2N)$ matrix given by
\beq
\label{defM0}
M_0=\,\begin{pmatrix}
{\rm Id}_{2N} \cr
- {\tilde R}^{-1}_0\,R_0
\end{pmatrix}
\eeq
\end{definition}

The matrix $M_0$ is defined in order to give:
\beq
\forall\,\, e\in{\cal E}(T)\,\,, \qquad d\theta_e = \sum_{e'\in \calE_0} (M_0)_{e,e'} d\theta_{e'}.
\eeq

\begin{definition}
Finally, we define the restriction of the matrix $E$ to the basis $\mathcal{E}_0$ simply as the $(2N)\times(2N)$ matrix
\begin{equation}
\label{defE0}
E_0=P_0^T E\, P_0
\end{equation}
\end{definition}

We now have some simple results

\begin{lemma}
\label{REzero}
\begin{equation}
R\cdot E =0
\end{equation}
\end{lemma}
\proof{Trivial, just compute all cases.}

\begin{lemma}
\label{LdetE0}
\begin{equation}
\label{detE0one}
\det(E_0)=1
\end{equation}
\end{lemma}

\proof{
Remember that the basis $\calE_0$ is the set of edges of a union of 2 trees $T_0^*$ on the dual graph of the triangulation $T$, with $2N$ edges.
We have:
\beq
\det_{e'\in \calE_0,\, e\in \calE_0} (E_{e',e}) = 
\sum_\sigma (-1)^{\sigma}\,\,\prod_e E_{e,\sigma(e)}
\eeq
The product $\prod_e E_{e,\sigma(e)}$ is non-vanishing only if $e$ and $\sigma(e)$ are neighbors, and in particular, since $E_{e,e}=0$, we must have $\sigma(e)\neq e$.
First observe that a cycle of $\sigma$ corresponds to a vertex of $T_0^*$, i.e. $\sigma$ can have only cycles of lengths $2$ or $3$.

If $\sigma$ contains a cycle of length $3$, $\sigma(v)=v'$, $\sigma(v')=v''$ and $\sigma(v'')=v$, we have a factor $E_{v,v'}E_{v',v''}E_{v'',v}$. Let $\sigma'$ be the permutation obtained from $\sigma$ by reversing the cycle $v\to v'\to v''\to v$ into $v\to v''\to v'\to v$, we have
$E_{v,v'}E_{v',v''}E_{v'',v} = -E_{v,v''}E_{v'',v'}E_{v',v}$, and therefore the contribution of $\sigma'$ cancels that of $\sigma$.

This shows that only $\sigma$'s which have only cycles of length $2$ contribute, i.e. $\sigma=$product of transpositions, and $(-1)^\sigma=(-1)^N$. $\sigma$ is thus an involution without fixed points, and can be represented as a perfect matching. Notice that $E_{e,e'}E_{e',e}=-1$, therefore $\prod_e E_{e,\sigma(e)}=(-1)^N$, finally we have:
\beq
\det_{e'\in \calE_0,\, e\in \calE_0} (E_{e',e}) = (-1)^N\,(-1)^N\#{\rm perfect\,matchings\,on}\,\calE_0
\eeq
Let us prove that there is only 1 perfect matching on a (at most trivalent) forest with $2N$ edges.
Start from an edge $e$ which is a leaf, and apply the following procedure:

1) it has 1 or 2 neighboring edges on $\calE_0^*$. 

2) If it has only 1 neighbor $e'$, we match $e$ with its unique neighbor $e'$, and we start  to 1) with the forest $\calE_0^*\setminus \{e,e'\}$.

3) If it has 2 neighbors $e'$ and $e''$, and since $\calE_0^*$ is at most trivalent, this means that removing $e$ and $e'$ (resp. $e$ and $e''$) disconnects the tree $\calE_0^*$ into two subtrees whose total number of edges is $2N-2$. If the subtrees obtained by removing $e'$ (resp. $e''$) are odd, then the subtrees obtained by removing $e''$ (resp. $e'$) are even. In other words there is one and only one possibility to match $e$ with $\sigma(e)=e'$ or $\sigma(e)=e''$  such that $\calE_0^*\setminus (e,\sigma(e))$ is the union of even subtrees.
Then, remove $e$ and $\sigma(e)$ from $\calE_0^*$, and start at 1) again with the  subtrees.

\medskip

In the end, we get the unique perfect matching on $\calE_0$.

This gives:
\beq
\det_{e'\in \calE_0,\, e\in \calE_0} (E_{e',e}) =1.
\eeq

}

\begin{lemma}
\label{LEME0M}
\beq
\label{EME0M}
E = M_0\,E_0\, M_0^T
\eeq
\end{lemma}

\proof{
Lemma \ref{REzero}, $R.E=0$ implies that $E$ is of the form 
\beq
E   \ = \ \begin{pmatrix}
E_0 & \tilde E_0 \cr
- \tilde R_0^{-1} R_0 E_0 & - \tilde R_0^{-1} R_0 \tilde E_0
\end{pmatrix}
\ =\  M_0\cdot (E_0, \tilde E_0)
\eeq
with $\tilde E_0$ some $(2N)\times (N+3)$ matrix.
And the condition that $E=-E^T$ implies that 
$\tilde R_0 \tilde E_0^T = R_0 E_0 $, and thus
\beq
E = M_0\,E_0\, M_0^T
\eeq
}
We now express the Jacobian matrix $J$ and the Kähler metric $D$ in terms of these matrices.

\begin{proposition}
\label{pJAE}
We have:
\beq
\label{JAE}
J_{v,e}={\partial \theta_e\over\partial z_v} = \frac{\ii}{2}\,\sum_{e'} A_{v,e'}\,E_{e',e}=\frac{\ii}{2}\,A\cdot E
\quad,\qquad
\bar J_{v,e}={\partial \theta_e\over\partial \bar z_v} = \frac{-\ii}{2}\,\sum_{e'} \bar A_{v,e'}\,E_{e',e}=\frac{\ii}{2}\,\bar A\cdot E
\eeq
\end{proposition}
This is nothing but \ref{JveExpl}.

\proof{
Let $e$ an edge, and $v,v'$ its two adjacent vertices.
We denote $(v,v',v'')$ and $(v,v''',v')$ the two positively oriented triangles adjacent to $e$, and $f,f'$ the center of their respective circumscribed circles.
Notice that in the triangle $(v,f,f')$ we have:
\beq
\theta_e=\pi-\alpha_{f,e}-\alpha_{f',e}
\eeq 
and since $f$ is the center of the circumscribed circle to $(v,v',v'')$ we have
\beq
\alpha_{f,e} = \widehat{vv''v'} = \Arg \frac{z_{v'}-z_{v''}}{z_{v}-z_{v''}} = \Im\,\ln{\left(\frac{z_{v'}-z_{v''}}{z_{v}-z_{v''}}\right)}
\eeq
and similarly
\beq
\alpha_{f',e} = \widehat{v'v'''v}= \Arg \frac{z_{v}-z_{v'''}}{z_{v'}-z_{v'''}} = \Im\,\ln{\left(\frac{z_{v}-z_{v'''}}{z_{v'}-z_{v'''}}\right)}.
\eeq 
That gives:
\bea
d\theta_e 
&=& \Im\,\,\left(\frac{dz_{v}-dz_{v''}}{z_{v}-z_{v''}} -  \frac{dz_{v'}-dz_{v''}}{z_{v'}-z_{v''}} + \frac{dz_{v'}-dz_{v'''}}{z_{v'}-z_{v'''}} - \frac{dz_{v}-dz_{v'''}}{z_{v}-z_{v'''}} \right) \cr
&=& \frac{-1}{2\ii}\,\sum_v \sum_{e'} \left( \frac{dz_v}{z_v-z_{v+e'}} E_{e',e} -  \frac{d\bar z_v}{\bar z_v-\bar z_{v+e'}} E_{e',e} \right)
\eea
}

\begin{proposition}
\label{pDAEA}
The Kähler matrix $D=(D_{u,\bar v}$ defined by \ref{Dkd2zA} can be rewritten as
\begin{equation}
\label{DAEAT}
D= \frac{1}{4\ii}\,A\cdot E\cdot A^\dagger
\end{equation}
\end{proposition}
\proof{
Writing for an oriented triangle $f=(v,v',v'')$:
\bea
-\,d \mathrm{Vol}(f) 
&=& d\alpha_v\,\,\ln{|z_{v'}-z_{v''}|} +d\alpha_{v'}\,\,\ln{|z_{v''}-z_{v}|} +d\alpha_{v''}\,\,\ln{|z_{v}-z_{v'}|}  \cr
&=& \frac{1}{2\ii} \Big( 
\left(\frac{dz_{v'}-dz_{v}}{z_{v'}-z_{v}}-\frac{dz_{v''}-dz_{v}}{z_{v''}-z_{v}}\right)\,\ln{|z_{v'}-z_{v''}|} - {\rm c.c.} \cr
&& + \left(\frac{dz_{v''}-dz_{v'}}{z_{v''}-z_{v'}}-\frac{dz_{v}-dz_{v'}}{z_{v}-z_{v'}}\right)\,\ln{|z_{v''}-z_{v}|} - {\rm c.c.} \cr
&& + \left(\frac{dz_{v}-dz_{v''}}{z_{v}-z_{v''}}-\frac{dz_{v'}-dz_{v''}}{z_{v'}-z_{v''}}\right)\,\ln{|z_{v}-z_{v'}|} - {\rm c.c.} \Big)\cr
\eea
i.e.
\bea
-2\ii\,\frac{\partial \mathrm{Vol}(f)}{\partial z_v}
&=& \frac{1}{z_v-z_{v'}}\,\ln{|z_{v'}-z_{v''}|}-\frac{1}{z_v-z_{v''}}\,\ln{|z_{v'}-z_{v''}|} \cr
&& - \frac{1}{z_v-z_{v'}}\,\ln{|z_{v}-z_{v''}|} +\frac{1}{z_v-z_{v''}}\,\ln{|z_{v}-z_{v'}|}  \cr
&=& \sum_{e,e'} A_{v,e}\,E_{e,e'}\,\ln{ l_{e'}}
\eea
the $\bar\partial $ act only on the log terms, and give for an edge $e=(v,v')$:
\beq
\frac{\partial \ln{l_{e}}}{\partial \bar z_v} = \frac{1}{2}\,\bar A_{v,e}.
\eeq
}

Finally one has the following result. Albeit simple, it is very important, so let us promote it to a theorem
\begin{theorem}
\label{tAEAT0}
We have the identity
\begin{equation}
\label{eAEAT0}
A\cdot E\cdot A^T=0
\end{equation}
\end{theorem}
\proof{
If $v=v'$, let $v_i$ be the neighboring vertices of $v$, oriented positively, we have:
\beq
(AEA^t)_{v,v} 
= \sum_i \frac{1}{z_v-z_{v_i}}\,\left(\frac{1}{z_{v}-z_{v_{i-1}}}-\frac{1}{z_{v}-z_{v_{i+1}}}\right)
=0 
\eeq
If $v\neq v'$ are neighbors, and the (positively oriented) triangles adjacent to the edge $(v,v')$ are denoted $(v,v',v'')$ and $(v,v''',v')$, we have
\bea
(AEA^t)_{v,v'} 
&=& \frac{1}{z_v-z_{v'}}\,\left(\frac{1}{z_{v'}-z_{v''}}-\frac{1}{z_{v'}-z_{v'''}}\right) \cr
&& + \frac{1}{z_v-z_{v''}}\,\left(\frac{1}{z_{v'}-z_{v}}-\frac{1}{z_{v'}-z_{v''}}\right) \cr
&& + \frac{1}{z_v-z_{v'''}}\,\left(\frac{1}{z_{v'}-z_{v'''}}-\frac{1}{z_{v'}-z_{v}}\right) \cr
&=& 0
\eea
Any other element of $(A E A^t)_{v,v'}=0$.
}

\subsection{Proof of theorem \ref{ThDas3T}}
\label{proofDas3T}
Let us choose a face $f_0$, dual to a triangle $(v_1,v_2,v_3)$. We want to compute the determinant in \ref{defDJacob}
\begin{equation}
\label{Jtilde}
\td J=\det\left( J_T(z)_{\setminus\{1,2,3\}\times {\mathcal{E}}_0}\right)
\end{equation}
Using Prop.~\ref{pJAE} we have:
\beq
\label{Jtexpl}
\td J = (\ii/2)^{N} \,(-\ii/2)^N\,\, \det_{v\in {\cal V}(T)\setminus\{v_1,v_2,v_3\}, e\in \calE_0} \left( \begin{array}{l}
\sum_{e'\in E} A_{v,e'}\,E_{e',e} \cr
\sum_{e'\in E} \bar A_{v,e'}\,E_{e',e} \cr
\end{array}\right)
\eeq
The Cauchy-Binet identity gives
\beq
4^N\,\td J = \sum_{I\subset E,\, \# I=2N}\,\,
\det_{v\in {\cal V}(T)\setminus\{v_1,v_2,v_3\}, e'\in I} \left( \begin{array}{l}
\sum_{e'\in I} A_{v,e'} \cr
\sum_{e'\in I} \bar A_{v,e'} \cr
\end{array}\right)
\,\, \times \,\,
\det_{e'\in I,\, e\in \calE_0} (E_{e',e})
\eeq
The first determinant can be written as a sum over  permutations ${\cal V}(T)\setminus\{v_1,v_2,v_3\} \cup {\cal V}(T)\setminus\{v_1,v_2,v_3\} \to I$, which we decompose into two applications:
\beq
\begin{array}{l r  c l r}
\sigma: & {\cal V}(T)\setminus\{v_1,v_2,v_3\} \to I
& \qquad\qquad & \bar \sigma : & {\cal V}(T)\setminus\{v_1,v_2,v_3\} \to I \cr
& v\mapsto e & \qquad \qquad & & v\mapsto e
\end{array}
\eeq
and we call $I'\subset I$ the image of $\sigma$ and $\bar I'\subset I$ the image of $\bar\sigma$, they have to be disjoint $I'\cap \bar I'=\emptyset$. We thus write:
\beq
4^N\,\td J = \sum_{I\subset {\cal E}(T),\,\#I=2N}
\sum_{\sigma\cup \bar\sigma: {\cal V}(T)\setminus\{v_1,v_2,v_3\} \to I} (-1)^{(\sigma\cup \bar\sigma)} \prod_{v\neq v_1,v_2,v_3} A_{v,\sigma(v)}\,\bar A_{v,\bar\sigma(v)}\,\,\quad \det_{e'\in I,\, e\in \calE_0} (E_{e',e})
\eeq
Since $A_{v,e}$ is non-zero only if $e$ is adjacent to $v$, we have the further constraint that $\sigma(v)$ (resp. $\bar\sigma(v)$) be a neighbouring edge to $v$.

\begin{figure}[h!]
\begin{center}
\includegraphics[width=7.cm]{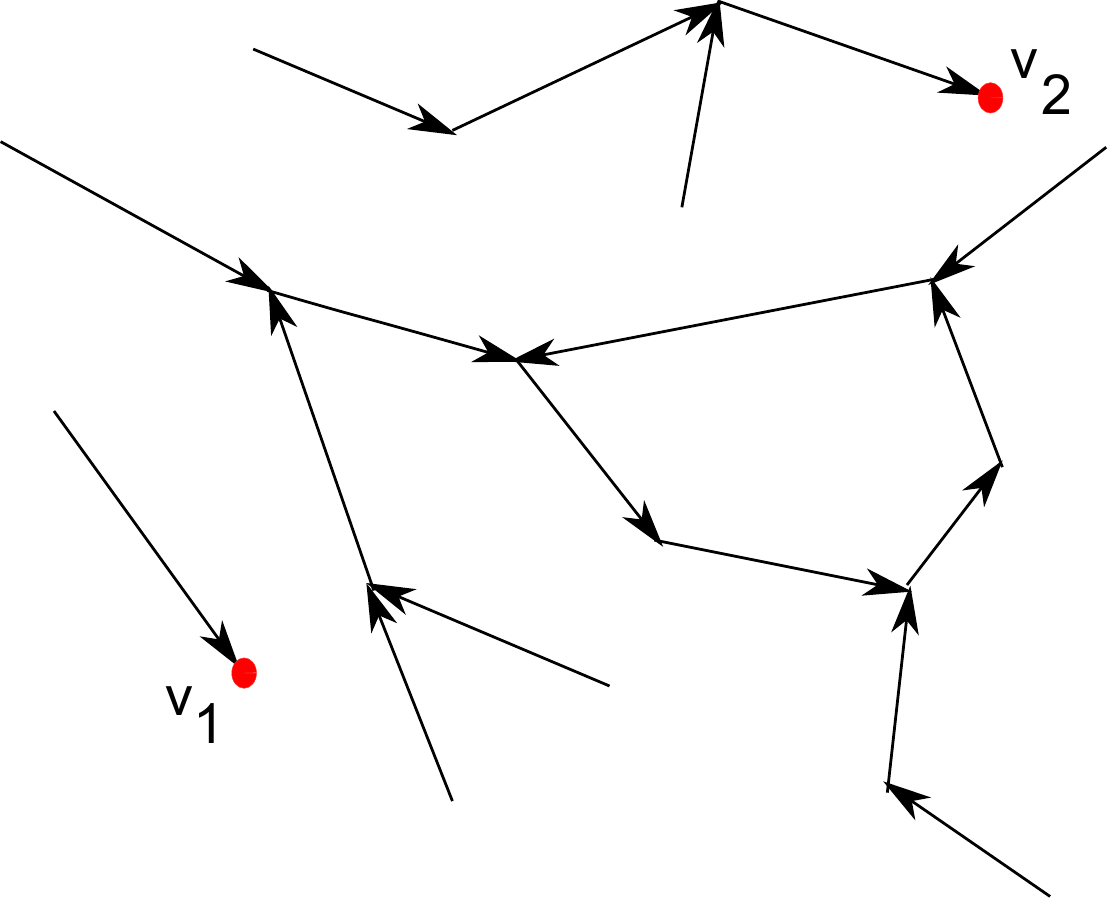}
\caption{A map $\sigma:v\mapsto e$ maping a vertex to an adjacent edge, can be encoded as a graph. Each pair $(v,\sigma(v))$ is represented by an arrow on edge $\sigma(v)$ originating at $v$. The graph is thus such that each vertex has exactly one outgoing arrowed edge. It may have any number of incoming edges. The vertices at the leaves of the graph can only be $v_1, v_2$ or $v_3$ since every other vertex has an outgoing arrow. The only possibility of not ending on $v_1,v_2,v_3$, is to end on a loop. The graph is thus a union (forest) of trees and 1-loop trees.}
\label{figtreeI1}
\end{center}
\end{figure}

One can thus represent a pair $(v,\sigma(v))$ (resp. $(v,\bar\sigma(v))$) as a white (resp. black) arrow on edge $e=\sigma(v)$ (resp. $\bar\sigma(v)$) with origin at $v$.
Every vertex (except $v_1,v_2,v_3$) must have a unique white (resp. black) outgoing arrow. 
Therefore, the set of white arrows $I'=\sigma({\cal V}(T)\setminus\{v_1,v_2,v_3\})$ (resp. $\bar I'=\bar\sigma({\cal V}(T)\setminus\{v_1,v_2,v_3\})$) is a forest of  trees with 0 or 1 loop, see fig.\ref{figtreeI1}. 

Notice that trees with no loop must end on $v_1, v_2$ or $v_3$ (because any other vertex would have an outgoing arrow).
Let us assume that it has $k_0$ trees connected components, and $k_1$ 1-loop trees connected components. Since a tree has one more vertex than edges, and 1-loop trees have as many vertices as edges, and since we have $N$ edges, the number of vertices is thus $N+k_0$.
Since $T$ has $N+3$ vertices (including $v_1,v_2,v_3$) we get that $k_0\leq 3$, and $k_0>0$ implies that $k_0$ of the points $v_1,v_2,v_3$ are reached. But notice that the vertices $v_1,v_2,v_3$ can't be reached on a 1-loop tree. If we add to $I'$ the 3 edges of the triangle $(v_1,v_2,v_3)$ (they form a loop), we see that $I'\cup ((v_1,v_2),(v_2,v_3),(v_3,v_1)$ is a forest of 1-loop trees only.

\begin{figure}[h!]
\begin{center}
\includegraphics[width=7.cm]{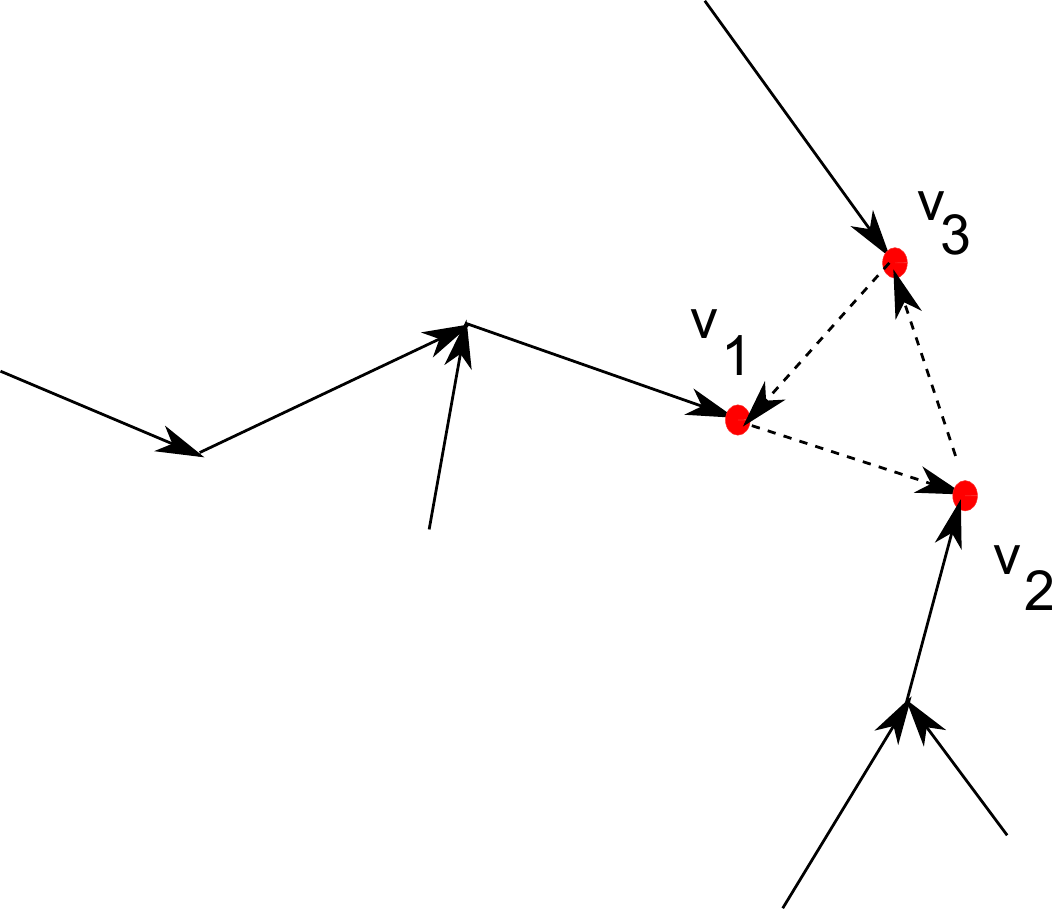}
\caption{The image $I'$ of $\sigma$ (resp. $\bar I'$ image of $\bar\sigma$)  must be a union of 3 disjoint trees (possibly with no edges) ending at $v_1,v_2,v_3$, i.e. if we add the edges of the triangle $(v_1,v_2,v_3)$ we must have a 1-loop tree whose unique loop is the triangle $(v_1,v_2,v_3)$.}
\label{figtreeI3}
\end{center}
\end{figure}

Now, Imagine that $I'$ contains one (or more) loop different from the triangle $(v_1,v_2,v_3)$. One can then find another map $\tilde\sigma:{\cal V}(T)\setminus\{v_1,v_2,v_3\})\to I'$ obtained by reversing the orientation of the loop. Its signature is multiplied by $(-1)^{{\rm length\,}-1}$, whereas the product $\prod_v A_{v,\sigma(v)}$ gets multiplied by $(-1)^{{\rm length\,}}$, i.e. the contribution of $\td\sigma$ is the opposite of that of $\sigma$, and the sum of the contributions of $\sigma$ and $\td\sigma$ cancel.
In the end we have to consider only $I'$ without loops other than $(v_1,v_2,v_3)$.
Therefore $I'$ must be a union of 3 trees (possibly empty) ending respectively on $v_1,v_2,v_3$.
Cf fig.\ref{figtreeI3}. We have the same for $\bar I'$.

Moreover, because of the factor $\det_{e'\in I,\, e\in \calE_0} (E_{e',e})$ which vanishes when $I$ is not a basis, we have an extra requirement on $I'$ and $\bar I'$, which  is that  $I=I'\cup \bar I'$ must be a basis, i.e. ${\cal E}(T)\setminus I$ is a 1-loop tree, with a loop of odd length. Since the edges of the triangle $(v_1,v_2,v_3)$ are not contained in $I'$ nor $\bar I'$, they have to be in ${\cal E}(T)\setminus I$, and thus the unique loop of ${\cal E}(T)\setminus I$ must be the triangle $(v_1,v_2,v_3)$.

Eventually this gives
\beq
4^N\,\td J = \sum_{{\rm trees}\, I',\bar I'}\,\,
\prod_v A_{v,\sigma(v)}\,\bar A_{v,\bar\sigma(v)}\,\,\quad \left((-1)^{(\sigma,\bar\sigma)}\,\, \det_{e'\in I'\cup\bar I',\, e\in \calE_0} (E_{e',e})\right)
\eeq
where the sum is over disjoint sets $I'$, $\bar I'$ of $N$ edges forming trees rooted at $v_1,v_2,v_3$, and such that the complementary set ${\cal E}(T)\setminus (I'\cup\bar I')$ is a 1-loop tree whose loop is the triangle $(v_1,v_2,v_3)$.

In the end we have:

\beq
\td J = {1\over 4^N} \sum_{{\rm trees}\, I',\bar I'}\,\,
 (\pm 1) \prod_{(v,e)\in I'} A_{v,e}\,\,\prod_{(v,e)\in\bar I'}\bar A_{v,e}\,
\eeq
where the sum is over disjoint sets $I'$, $\bar I'$ of $N$ edges forming oriented trees ending at $v_1,v_2,v_3$, and such that the complementary set ${\cal E}(T)\setminus (I'\cup\bar I')$ is a 1-loop tree whose loop is the triangle $(v_1,v_2,v_3)$.
This implies \ref{defDJacob}, i.e. Th.~\ref{ThDas3T}.

In particular, the sign factor $(\pm1)=\epsilon(\mathcal{F})$ in \ref{defDJacob} is given explicitly by
\beq
\label{Fexpl}
\epsilon(\mathcal{F}) = (-1)^{(\sigma,\bar\sigma)} \det_{e'\in I'\cup\bar I',\, e\in \calE_0} (E_{e',e})
\eeq

Remark:
The 3 sets $I'$, $\bar I'$, and $I''={\cal E}(T)\setminus (I'\cup \bar I'\cup \{(v_1,v_2),(v_2,v_3),(v_3,v_1)\})$ play equivalent roles, the 3 of them have cardinal $N$ and are made of 3 trees ending on $v_1,v_2,v_3$, and are pairwise disjoint.

Remark that exchanging $I'$ and $\bar I'$ exchanges the $z_v$'s and the $\bar z_v$'s, and changes the prefactor by a $(-1)^N$.
\beq
\td J=-\bar{\td J}.
\eeq
This is Theorem \ref{TDasKha}.

\subsection{Proof of theorem \ref{ThConvD}}
\label{proofThConvD}
\begin{figure}[h]
\begin{center}
\includegraphics[width=2in]{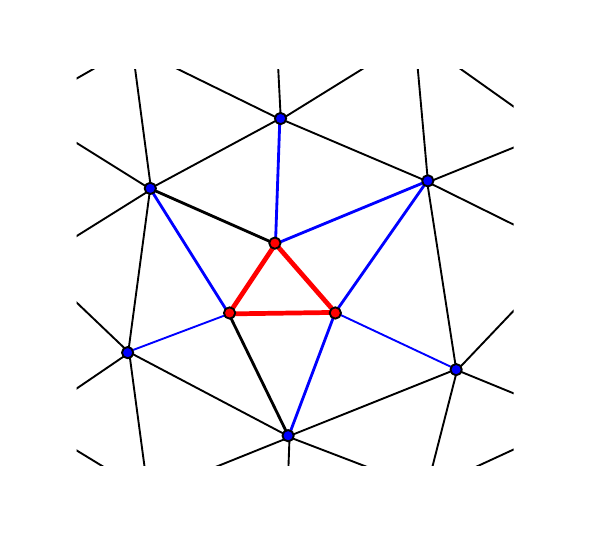}
\includegraphics[width=2in]{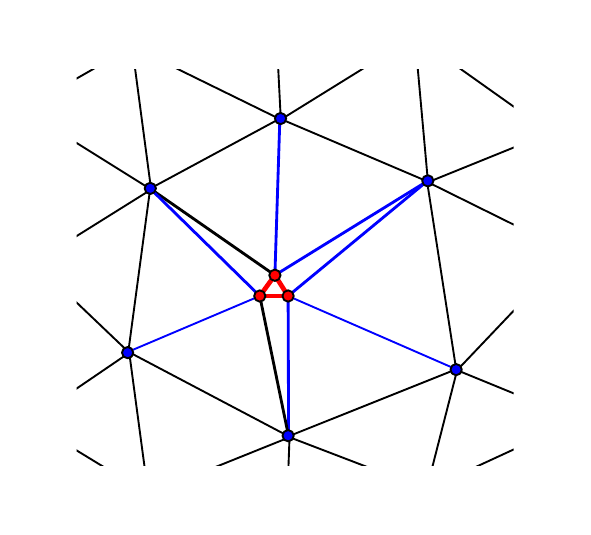}
\includegraphics[width=2in]{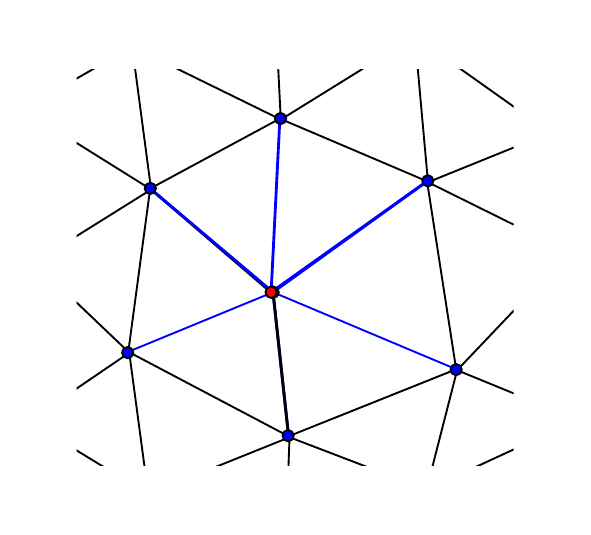}
\caption{Collapse of a triangulation $T$ to the triangulation $T_0$ as $x\to 0$}
\label{fScaling}
\end{center}
\end{figure}

\begin{figure}[h]
\begin{center}
\includegraphics[width=2in]{scaling-1}
\includegraphics[width=2in]{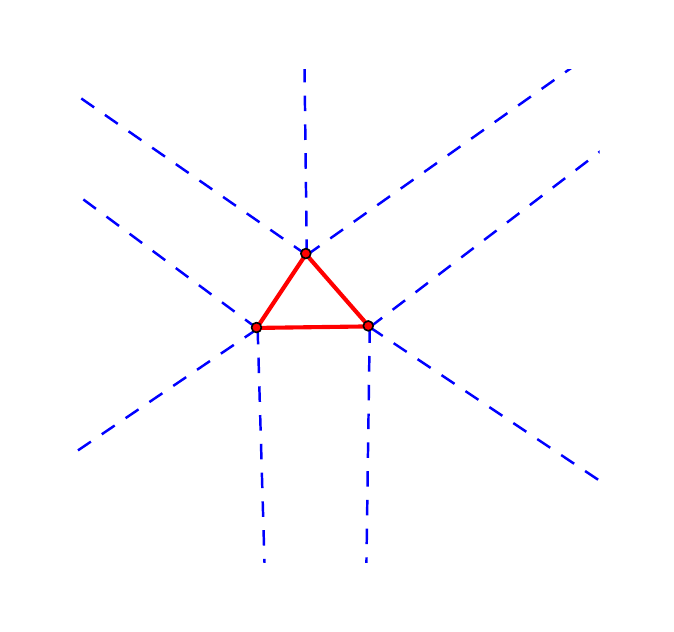}
\caption{Inverse scaling of a triangulation $T$ to the triangulation $\tilde T_0$ as $x\to 0$. The $\tilde{\tilde{T}}_0$ part is in red.}
\label{fInvScaling}
\end{center}
\end{figure}

To prove Th.~\ref{ThConvD}, let us consider how the term associated to a given 3-tree $\mathcal{F}$ scales when a subset of vertices $\{z_v,\,v\in\mathcal{V}_0\subset\{1,N+3\}$ collapse to a single point $Z_0$. For simplicity we consider the case where $\mathcal{V}_0$ does not contain any of the fixed vertices $\{1,2,3\}$. So let us rescale 
\begin{equation}
\label{zRescaling}
z_v\to z_v(x)=Z_0+x\, (z_v-Z_0)
\qquad \text{if}\quad v\in\mathcal{sV}_0
\end{equation}
and study how the measure \ref{dmuF}
\begin{equation}
\label{dmuFx}
d\mu_\mathcal{F}(z)= \prod_{v\notin\triangle} d^2 z_v\  \prod_{\vec e=(v\to v')\in\,\mathcal{I}} {1\over z_v-z_{v'}}
 \times
 \prod_{\vec e=(v\to v')\in\,\mathcal{I}'} {1\over \bar z_v-\bar z_{v'}} 
\end{equation}
scales as $x\to 0$. It is clear that for $x\in{\mathbb{R}}^{*}_+$ small enough and the original $z_v$ fixed, the topology of the Delaunay triangulation should not depend on $x$, and it should collapse as $x\to 0$ from a triangulation $T$ with $N+3$ vertices to the triangulation $T_0$ with $N+2-P$ vertices (where $P=|\mathcal{V}_0|$), such that the $P$ vertices of $\mathcal{V}_0$ have been replaced by the vertex $Z_0$.
This is depicted in Fig.~\ref{fScaling}.
Similarily, if we keep the vertices of $\mathcal{V}_0$ fixed and send the other vertices of $T$ to infinity via the inverse scaling
\begin{equation}
\label{ }
z_v\to \tilde z_v(x)=Z_0+ x^{-1}\, (z_v-Z_0)
\qquad \text{if}\quad v\notin\mathcal{sV}_0
\end{equation}
we obtain a Delaunay triangulation $\tilde T_0$ with $P+1$ vertices: the $P$ vertices on $\mathcal{V}_0$ and an additional vertex et $\infty$.
This is depicted in Fig.~\ref{fInvScaling}.

It is clear that the most singular terms will come from the 3-trees $\mathcal{F}=(\mathcal{I},\mathcal{I}',\mathcal{I}'')$ such that the restrictions $\mathcal{I}_0$ and $\mathcal{I}'_0$ of the trees $\mathcal{I}$ and $\mathcal{I}'$ to the triangulation $\tilde{\tilde T}_{0}$ (obtained from $\tilde T_0$ by removing the vertex at $\infty$) have the maximal number of lines.
Since $\mathcal{I}$, $\mathcal{I}'$ and $\mathcal{I}''$ are disjoint, $\mathcal{I}_0$, $\mathcal{I}'_0$ and $\mathcal{I}''_0$ are also disjoint, hence are disjoint  spanning forests of $\tilde{\tilde T}_{0}$, whose union is the set of links of $\tilde{\tilde T}_{0}$.
From this it follows that
\begin{align}
\label{}
\#\ \text{vertex}\ \tilde{\tilde T}_{0}&= \#\ \text{links}\ \mathcal{I}_0 +   \#\ \text{connected components}\ \mathcal{I}''_0    \nonumber\\
&=\#\ \text{links}\ \mathcal{I}'_0 +   \#\ \text{connected components}\ \mathcal{I}'_0    \nonumber\\
    & = \#\ \text{links}\ \mathcal{I}''_0 +   \#\ \text{connected components}\ \mathcal{I}''_0 
\end{align}
and 
\begin{equation}
\label{ }
\#\ \text{links}\ \mathcal{I}_0 +\#\ \text{links}\ \mathcal{I}'_0 + \#\ \text{links}\ \mathcal{I}''_0 = \#\ \text{links}\ \tilde{\tilde T}_{0}
\end{equation}
since $\tilde{\tilde T}_{0}$ has the topology of a disk, its boundary (or hull) is a circle and is such that
\begin{align}
\label{}
  \#\ \text{links of the hull of}\ \tilde{\tilde T}_{0}  =   \#\ \text{vertex of the hull of}\ \tilde{\tilde T}_{0} 
\end{align}
From Euler formula and the fact the $\tilde{\tilde T}_{0} $ is a triangulation of the disk one has
\begin{equation}
\label{ }
 \#\ \text{vertex}\ \tilde{\tilde T}_{0} - \#\ \text{links}\ \tilde{\tilde T}_{0} +  \#\ \text{triangles}\ \tilde{\tilde T}_{0} =1
\end{equation}
and
\begin{equation}
\label{ }
2\  \#\ \text{links}\ \tilde{\tilde T}_{0} - \#\ \text{links of the hull of}\ \tilde{\tilde T}_{0} =  3\  \#\ \text{vertex}\ \tilde{\tilde T}_{0}
\end{equation}
Each connected component of the spanning forests has at least one point on the hull, hence
\begin{equation}
\label{cclevh}
 \#\ \text{connected components of}\  \mathcal{I}_0,\  \mathcal{I}'_0\ \text{and}\ \mathcal{I}''_0 \le  \#\ \text{vertex of the hull of}\ \tilde{\tilde T}_{0}
\end{equation}
Combining these relations we get
\begin{equation}
\label{ineqliver}
\#\ \text{links}\ \mathcal{I}_0 +\#\ \text{links}\ \mathcal{I}'_0 \ \le\  2\  \#\ \text{vertex}\ \tilde{\tilde T}_{0}-3
\end{equation}
Now under the rescaling \ref{zRescaling} the measure $d\mu_{\mathcal{F}}$ scales as
\begin{equation}
\label{ }
d\mu_{\mathcal{F}}(z(x))\ \propto  \ x^n
\quad,\qquad n=\ 2\  \#\ \text{vertex}\ \tilde{\tilde T}_{0}-(\#\ \text{links}\ \mathcal{I}_0 +\#\ \text{links}\ \mathcal{I}'_0)
\end{equation}
and \ref{ineqliver} implies that
\begin{equation}
\label{ }
n>2
\end{equation}
This ensures global convergence of the integral over the measure when all the points of $\mathcal{V}_0$ tend towards $Z_0$ at the same rate.
In the language of renormalization theory, the ``superficial degree of convergence" of the subgraph $\mathcal{V}_0$, given by $n-2$, is strictly positive.

The analysis of the subdivergences when subsets of points collapse at different rates is more involved but goes along the same line. The decomposition in forests of the total measure $d\mu(z)$ given by \ref{Das3trees} ensures that the analysis by power counting is valid and that the measure is absolutely convergent.

One should note that the inequality \ref{ineqliver} is saturated when the inequality \ref{cclevh} is saturated for the spanning forest $\mathcal{I}''_0$, i.e. when
\begin{equation}
\label{ }
 \#\ \text{connected components of}\  \mathcal{I}''_0\ =\   \#\ \text{vertex of the hull of}\ \tilde{\tilde T}_{0}
\end{equation}
which implies
\begin{equation}
\label{ }
 \#\ \text{connected components of}\  \mathcal{I}_0\ +\  \#\ \text{connected components of}\  \mathcal{I}'_0\ =\ 3
 \end{equation}
This probably implies some factorization property for the measure.

\subsection{Proof of theorem \ref{TDasKha}}
\label{proofDasKha}
\def\Qzero{Q_{\scriptscriptstyle{0}}}
Let us denote by $\Qzero$ the $N\times (N+3)$ projection matrix obtained by projecting out the first three fixed vertices
\def\Qzero{Q_{\scriptscriptstyle{0}}}
\begin{equation}
\label{ }
\Qzero=(\delta_{u,v})_{\begin{smallmatrix}u=4,N+3\\v=1,N+3\end{smallmatrix}}
\end{equation}
So that we project out the first 3 lines of the the matrix $A$ and the first 3 lines and columns of the matrix $D$ by
\begin{equation}
\label{ }
A'=\Qzero A \quad,\qquad D'=\Qzero D \Qzero^T
\end{equation}
Using Prop.~\ref{pDAEA}  and Th.~\ref{tAEAT0} we have
\beq
\begin{pmatrix} A' \cr  -\bar A' \end{pmatrix} E \begin{pmatrix} A' \cr  -\bar A' \end{pmatrix}^T
= \begin{pmatrix} 0 &  -4\ii D' \cr  -4\ii\bar D'& 0 \end{pmatrix}
\eeq
so that
\begin{equation}
\label{thefirstD}
\left| { \det\left(  \begin{pmatrix} A' \cr - \bar A' \end{pmatrix} E \begin{pmatrix} A' \cr - \bar A' \end{pmatrix}^T  \right) } \right| =
4^{2N} \det(D')^2
\end{equation}
Using that $E=M_0 E_0 M_0^T$ (Lemma~\ref{LEME0M}, eq.~\ref{EME0M}) we have
\beq
\det\left(\begin{pmatrix} A' \cr - \bar A' \end{pmatrix} E \begin{pmatrix} A' \cr - \bar A' \end{pmatrix}^T\right)
= \det\left(\begin{pmatrix} A' \cr - \bar A' \end{pmatrix} M_0 E_0 M_0^T \begin{pmatrix} A' \cr - \bar A' \end{pmatrix}^T\right)
\eeq
Since $E_0$ is an invertible matrix and $\det E_0=1$ (Lemma~\ref{LdetE0})  we can write:
\beq
\det\left(\begin{pmatrix} A' \cr - \bar A' \end{pmatrix} E \begin{pmatrix} A' \cr - \bar A' \end{pmatrix}^T\right)
= \det\left(\begin{pmatrix} A' \cr - \bar A' \end{pmatrix} M_0 E_0\right)\ 
 \det\left( E_0 M_0^T \begin{pmatrix} A' \cr - \bar A' \end{pmatrix}^T\right)
\eeq
and since $P_{0}^T M_0 =  1$ we have (using  Lemma~\ref{LEME0M}, eq.~ \ref{EME0M} )  
\begin{align}
\label{}
\det\left(\begin{pmatrix} A' \cr - \bar A' \end{pmatrix} E \begin{pmatrix} A' \cr - \bar A' \end{pmatrix}^T\right)    &
= \det\left(\begin{pmatrix} A' \cr - \bar A' \end{pmatrix} M_0 E_0 M_0^T P_{0}\right) \ 
 \det\left( P_{0}^T M_0 E_0 M_0^T \begin{pmatrix} A' \cr - \bar A' \end{pmatrix}^T\right)   \\
    & =  
    \det\left(\begin{pmatrix} A' \cr - \bar A' \end{pmatrix} E P_{0}\right)\ 
    \det\left( P_{0}^T E \begin{pmatrix} A' \cr - \bar A' \end{pmatrix}^T\right) 
\end{align}
One recognizes the determinant $\tilde J$ of the jacobian matrix defined in \ref{Jtilde} and \ref{Jtexpl} 
since 
\begin{equation}
\label{thefirstJ}
\tilde J ={1\over 4^N}\det\left(\begin{pmatrix} A' \cr - \bar A' \end{pmatrix} E P_{0}\right)
\end{equation}
\ref{thefirstD} and \ref{thefirstJ} give the final result
\begin{equation}
\label{dD2tJ2}
\det(D')^2={\tilde J}^2
\end{equation}
This gives Theorem~\ref{TDasKha}.

\subsection{Proof of proposition \ref{pHdens}}
We study in more details the conformal properties of the Kähler form $D$ and its determinant.
We consider a triangulation $T$ with $N+3$ points in the plane. No point is considered fixed, and the exterior triangle which contains the point at $\infty$ is taken to have a clock wise orientation, so that its area and its hyperbolic volumes are taken with a negative sign.
We have seen that $\mathcal{A}$ is minus the sum of the hyperbolic volumes associated to the triangles (the faces) of the  triangulation $T$ associated to the position of the vertices $z_1,z_{N+4}$.
\begin{equation}
\label{ }
\mathcal{A}=- \sum_{\mathrm{triangles}}\mathrm{Vol}(z_i,z_j,z_k)
\end{equation}
and the $(N+3)\times (N+3)$ Kähler matrix $D$ is
\begin{equation}
\label{ }
D_{i\bar\jmath}={\partial\over \partial z_i}{\partial\over \partial \bar z_{j}}\mathcal{A}
\end{equation}
A convenient basis to study the conformal properties of the matrix $D$ is to use the vectors 
\begin{equation}
\label{ }
\psi_a=
\begin{pmatrix}z_1^{a-1}&\ldots&z_{\scriptscriptstyle{N+3}}^{a-1}\end{pmatrix}
\quad,\qquad
\bar\psi_a=
\begin{pmatrix}
\bar z_1^{a-1} \\
\vdots\\
\bar z_{\scriptscriptstyle{N+3}}^{a-1}\end{pmatrix}
\quad,\qquad a=1,\cdots,N+3
\end{equation}
which define the $(N+3)\times(N+3)$ square matrices
\begin{equation}
\label{ }
\Psi=\begin{pmatrix}
\psi_1\\
\vdots\\
\psi_{\scriptscriptstyle{N+3}}
\end{pmatrix}
\quad,\qquad
\Psi^\dagger=
\begin{pmatrix}
\bar\psi_1&\ldots& \bar\psi_{\scriptscriptstyle{N+3}}
 \end{pmatrix}
\end{equation}
and to consider the $(N+3)\times (N+3)$ matrix
\begin{equation}
\label{ }
\tilde D=\Psi D \Psi^\dagger
\end{equation}
whose matrix elements are 
\begin{equation}
\label{ }
\tilde D_{ab}= z_i^{a-1}\, D_{i\bar\jmath}\,{\bar z}_j^{b-1}
\end{equation}

Since the hyperbolic volumes of ideal tetraedra Vol$(z_i,z_j,z_k,z_l)$ are invariant under global SL($2,\mathbb{C}$) conformal transformations
\begin{equation}
\label{ }
z\to w={a z+b\over c z+d}
\quad\text{i.e}\quad\mathrm{Vol}(z_i,z_j,z_k,z_l)=\mathrm{Vol}(w_i,w_j,w_k,w_l)
\end{equation}
it is clear that the matrix $D$ transform covariantly under global conformal transformations.
In particular,  the three generators of global conformal transformations correspond to the three zero modes of $D$
\begin{equation}
\label{ }
\psi_a D=0\quad\text{and}\quad D\bar\psi_a=0\quad\text{for}\ a=1,2,3
\end{equation}
This implies that the first three lines and three columns of the matrix $\tilde D$ are zero, so that $\tilde D$ takes the block form
\begin{equation}
\label{ }
\tilde D=\begin{pmatrix}
    0  &  0  \\
     0 &  D_0
\end{pmatrix}
\end{equation}
where $D_0=\left(\tilde D_{ab}\right)_{a,b>3}$ is a non-degenerate Hermitian $N\times N$ matrix.
Let us define the non-zero determinant of $\tilde D$ as
\begin{equation}
\label{ }
{\det}'(\tilde D)=\det(D_0)
\end{equation}
We want to express the relation between ${\det}'(\tilde D)$ and the determinant $\mathcal{D}_{\setminus{1,2,3}} =\det\left( D_{\backslash_{\scriptscriptstyle{1,2,3}}} \right)$ obtained by removing the first 3 lines and columns of $D$ (i.e. fixing the 3 points $(1,2,3)$).
Using the fact that the Jacobian of the change of variables given by the matrix $\Psi$ is nothing but the Vandermonde determinant
\begin{equation}
\label{ }
\det(\Psi)=\Delta_{\scriptscriptstyle{N+3}}(z_1,\cdots z_{\scriptscriptstyle{N+3}})=\prod_{i<j} (z_i-z_j)
\end{equation}
it is easy to show that the original determinant $\mathcal{D}_{\setminus{1,2,3}} $ is related to ${\det}'(D)$ by
\begin{equation}
\label{ }
\det\left( D_{\backslash_{\scriptscriptstyle{1,2,3}}} \right)=
\left|{\Delta_3(z_1,z_2,z_3)\over\Delta_{\scriptscriptstyle{N+3}}(z_1,\cdots z_{\scriptscriptstyle{N+3}})}\right|^2{\det} '(\tilde D)
\end{equation}
where $\Delta_3(z_1,z_2,z_3)$ is the Vandermonde determinant for the three fixed points $(z_1,z_2,z_3)$
\begin{equation}
\label{ }
\Delta_3(z_1,z_2,z_3)=(z_1-z_2)(z_1-z_3)(z_2-z_3)
\end{equation}
Since ${\det} '(\tilde D)$ and $\Delta_{\scriptscriptstyle{N+3}}(z_1,\cdots z_{\scriptscriptstyle{N+3}})$ do not depend on the choice of the 3 points that are fixed, this leads to the relation between the $\mathcal{D}_{\setminus{abc}}$ for two different choices of triplets of fixed points (different ``gauge fixings'' for global conformal invariance)
\begin{equation}
\label{D123toDabc}
\det\left( D_{\backslash_{\scriptscriptstyle{a,b,c}}} \right) = 
\left|{\Delta_3(z_a,z_b,z_c)\over  \Delta_3(z_1,z_2,z_3)}\right|^2
\det\left( D_{\backslash_{\scriptscriptstyle{1,2,3}}} \right)
\end{equation}
Since the full matrix $D$ and the matrices $\Psi$ and $\Psi^\dagger$ transform simply under global SL(2,$\mathbb{C}$) transformations, 
one deduces also that it is the density function $H$ defined by $\ref{Hdens}$ as
\begin{equation}
\label{Hdensity}
H=
{ \det\left( D_{\backslash_{\scriptscriptstyle{a,b,c}}} \right)   \over \left|\Delta_3(z_a,z_b,z_c)\right|^2}
\ =\ 
{ {\det}'\left( D\right)   \over \left|\Delta_{\scriptscriptstyle{N+3}}(z_1,\cdots z_{\scriptscriptstyle{N+3}})\right|^2}
\end{equation}
transforms covariantly as
\begin{equation}
\label{ }
z\to w={az+b\over c z+d}
\quad,\qquad H(z)
= \prod_{i=1}^{N+3}   {| ad-bc| \over \left|c z_i+d  \right|^2 } H(w)
\end{equation}

\subsection{Proof of propositon \ref{pPhiDPsi}}
To prove \ref{PhiDPsi}, it is enough to start from the explicit form \ref{DfExpl1} for the matrix elements of $D(f)$ in terms of the angles $(\alpha_1,\alpha_2, \alpha_3)$ of the triangle and of its circumcircle radius $R$ and to use the well known geometrical relations involving the lengths of the edges $(l_1,l_2,l_3)$ and the area $A$ of the triangle.
More precisely, using \ref{PhiDPsi}  gives for the matrix element
\begin{equation}
\label{ }
D_{1\bar 2}={-1\over 4\ii}{z_3-z_2\over A}\ {A\over R^2}\ {1\over 4\ii}{\bar z_1-\bar z_3\over A}
=-{  l_1 l_2 \over 16 A R^2} \e^{\ii\alpha_3}
\end{equation}
Using 
\begin{equation}
\label{ }
{\sin(\alpha_1)\over l_1}={\sin(\alpha_2)\over l_2}={\sin(\alpha_3)\over l_3}={1\over 2 R}
\quad,\qquad
\mathrm{Area}={l_1l_2l_3\over 4 R}
\end{equation}
we get
\begin{equation}
\label{ }
D_{1\bar 2}=-{1\over 4 l_3 R} \e^{\ii\alpha_3}=-{1\over 8 R^2}{\e^{\ii\alpha_3}\over \sin(\alpha_3)}=-{1\over 8 R^2}(\cot(\alpha_3)+\ii)
\end{equation}
which is precisely \ref{DfExpl1}. The proof is similar for the diagonal elements.

\section{Conclusion}
In this article, we have investigated the relation between planar maps, considered as discretized combinatoric formulations of 2d gravity, and their planar embedding in the complex plane, in view of getting a better understanding of their relation with the conformal field theory and topological field theory formulations of 2d gravity.
We started from a natural extension of the circle packing and circle pattern methods, relying on Euclidean structures, i.e. triangulations with edge angle variables, and patterns of circumcircles of Delaunay triangulations. 
We have shown that  the uniform measure on random planar maps, equipped with the uniform Lebesgue measure on edge angles variables, gives a conformally invariant spatial point process on the complex plane or the Riemann sphere, with very interesting properties.
We obtained an explicit representation for this measure in term of geometrical objects (triangle rooted spanning 3-trees) on Delaunay triangulations. 
This measure is also the volume form of a Kähler metric on the space of Delaunay triangulations, whose prepotential has a simple geometrical interpretation in term of 3d hyperbolic geometry.
It can also be written as a "discrete Fadeev-Popov" determinant (involving  discrete 
complex derivative operators $\nabla,\bar\nabla$), very similar to Polyakov's conformal Fadeev-Popov determinant, thus establishing a new link between planar maps and 2d gravity.
It can as well be written as a combination of Chern classes, thus also establishing a link with topological 2d gravity.

In our opinion the results presented in this article are interesting, since they bring together different aspects of 2d gravity, treated from the combinatorial, conformal and topological points of view, and offer a new pathway to study 2d gravity.
Nevertheless we have discussed here only the discrete case, where the number of points $N_v$ (and more generally the local density of points in the plane) is finite. Of course the most interesting thing to do next is to define and study the continuum limit ($N_v\to\infty$, or equivalently the large distance scaling properties of this measure in the plane). This is required in order to get a better understanding of the various continuum limits of 2d gravity.

\section*{Acknowledgements}
F. D. thanks Michel Bauer and Philippe Di Francesco for their interest and many useful discussions. The work of B.E. is partly supported by the Quebec government FQRNT.

%

\end{document}